\documentclass[a4paper,oneside,fleqn,12pt]{article}
\usepackage{amsmath,amssymb,type1cm,bm,color,txfonts}

\usepackage{graphicx,psfrag,epsf}

\def\v#1{\mbox{\boldmath $#1$}}

\newcommand{\vvsp}{\vspace{15pt}}

\newcommand{\E}{\mathrm{E}}

\newcommand{\tr}{\mathrm{tr}}

\newcommand{\sgn}{\mathrm{sgn}}
\newcommand{\supp}{\mathrm{supp}}

\bmdefine{\theb}{\theta}
\bmdefine{\X}{X}
\bmdefine{\x}{x}
\bmdefine{\etab}{\eta} \bmdefine{\mub}{\mu} \bmdefine{\I}{I}

\newtheorem{thm}{Theorem}
\newtheorem{lem}{Lemma}
\newtheorem{cor}{Corollary}
\newtheorem{ass}{Assumption}
\newtheorem{prop}{Proposition}

\makeatletter
\def\section{\@startsection {section}{1}{\z@}{-3.5ex plus -1ex minus-.2ex}{2.3ex plus .2ex}{\large\bf}}
\makeatother

\makeatletter
\def\subsection{\@startsection {subsection}{1}{\z@}{-3.5ex plus -1ex minus-.2ex}{2.3ex plus .2ex}{\normalsize\bf}}
\makeatother

\begin{document}

\begin{center}

{\bf \large \vspace{-5cm} Macroeconomic Forecasting and Variable Selection with a \\ 
Very Large Number of Predictors: A Penalized Regression Approach}\\[2mm]

{\normalsize Yoshimasa Uematsu\footnote{
		Corresponding address: Yoshimasa Uematsu, 
		The Institute of Statistical Mathematics, 10-3 Midori-cho, Tachikawa,
		Tokyo 190-8562, Japan. E-mail: uematsu@ism.ac.jp.}} 

{\normalsize \sl The Institute of Statistical Mathematics}\\[2mm] 

{\normalsize Shinya Tanaka\footnote{Otaru University of Commerce, Department of Economics, 3-5-21 Midori, Otaru, Hokkaido 047-8501, Japan. E-mail: stanaka@res.otaru-uc.ac.jp.}} 

{\normalsize \sl Otaru University of Commerce}\\[2mm]

{\normalsize March 3, 2017}

\end{center}

\vspace{-10mm}
\begin{abstract}
This paper studies macroeconomic forecasting and variable selection using a folded-concave penalized 
regression with a very large number of predictors. The penalized regression approach leads to sparse estimates 
of the regression coefficients, and is applicable even if the dimensionality of the model is much larger than the sample size. 
The first half of the paper discusses the theoretical aspects of a folded-concave penalized regression 
when the model exhibits time series dependence. Specifically, we show the oracle inequality and the oracle property for 
ultrahigh-dimensional time-dependent regressors.
The latter half of the paper shows the validity of the penalized regression using two motivating empirical applications. The first forecasts 
U.S.\ GDP with the FRED-MD data using the MIDAS regression framework, where there are  more than 1000 covariates, 
while the sample size is at most 200. The second examines how well the penalized regression screens the hidden portfolio with around 40 stocks 
from more than 1800 potential stocks using NYSE stock price data. Both applications reveal that the penalized regression provides remarkable results 
in terms of forecasting performance and variable selection. 
\end{abstract}

\allowdisplaybreaks

\section{Introduction}

Recent advancements in macroeconomic data collection have led to an increased focus on 
high-dimensional time series analysis. A more efficient and precise analysis can thus be realized if we elicit information appropriately from a large number of explanatory variables. 
However, a higher-dimensional model does not necessarily yield better performance in terms of forecasting and parameter estimation; in fact, the performance varies depending on the dimensionality and which estimation method is considered. 
Without appropriate dimension reduction, performance may be poor owing to accumulated estimation losses from redundant or unimportant variables. 
After seminal papers on factor-based (diffusion index) forecasting, such as Stock and Watson (2002), this is now common tool for forecasting with large datasets.  
Specifically, Stock and Watson (2012) showed that factor-based forecasts have a good performance in comparison with existing forecasting methods, including autoregressive forecast, pretest methods, Bayesian model averaging, empirical Bayes, and bagging. They concluded that it seemed difficult to outperform a factor-based forecast without introducing  nonlinearity and/or time-varying  parameters to a forecast model.

In this paper, we tackle the high-dimensional forecasting and estimation problem from another theoretical and empirical points of view. 
We employ {\it sparse} modeling, which can allow for {\it ultrahigh} dimensionality, where 
the number of regressors diverges sub-exponentially. The unknown sparsity can be recovered using a {\it folded-concave penalized regression} to pursue both 
prediction efficiency and variable selection consistency. 
In particular, we consider penalties including the smoothly clipped absolute deviation (SCAD) penalty introduced by Fan and Li (2001), the minimax concave penalty 
(MCP) proposed by Zhang (2010) as well as the $\ell_1$-penalty (Lasso) proposed by Tibshirani (1996). 
Previous studies on macroeconomic forecasting using sparse modeling include Bai and Ng (2008), De Mol et al.\ (2008), Kock and Callot (2015), Marsilli (2014), and Nicholson et al. (2015), but basically their estimation strategies are limited to the $ \ell_1 $-penalty. Although the $ \ell_1 $-penalty is expected to perform well as do 
the SCAD and MCP theoretically as we see in a later section, this is often insufficient in terms of model selection consistency while the SCAD 
and MCP can have this desirable property.  Moreover, it is difficult to find a statistical theory of penalized regression estimators in time a series context.

In the first half of this paper, we provide the comprehensive theoretical properties of the penalized regression estimator under suitable conditions for macroeconometrics from the perspective of both prediction efficiency and variable selection consistency. In fact, the theoretical aspects have been explored by many recent works on statistics, including B\"ulmann and van de Geer (2011), Fan and Lv (2011), 
Fan and Lv (2013), and Loh and Wainwright (2014), as well as the references therein.  
However, the results of these studies are not sufficient for time series econometrics. 
We in this paper derive a non-asymptotic upper bound for the prediction loss called the {\it oracle inequality}. This ensures that the forecasting value is reliable and it is an optimal forecast in the asymptotic sense. 
Likewise, we also show the estimation precision of the regression coefficient and the model selection consistency,  known as the {\it oracle property}; that is, 
it selects the correct subset of predictors and estimates the non-zero coefficients as efficiently as would be possible if we knew which variables were irrelevant. The oracle property provides 
another insight into the modeling of the variable of interest. In this regard, models can be selected by information criteria, such as the AIC and BIC. These have become popular owing to their tractability, 
however, they are limited when dealing with high-dimensional models because they demand an exhaustive search over all submodels. In contrast, the SCAD-type penalized regression yields simultaneous 
estimation and model selection, even in the ultrahigh-dimensional case.

In the second half of the paper, we shed light on the validity of the penalized regression in macroeconometrics by introducing two empirical applications.
The first one focuses on the oracle inequality. We consider to forecast quarterly U.S.\ real GDP with a large number of monthly predictors using MIDAS (MIxed DAta Sampling) regression 
framework originally proposed by Ghysels et al.\ (2007). 
Since the total number of parameters is much larger than that of observations,  this situation 
should be treated as an ultra-high dimensional problem. 
In contrast to the original MIDAS model of Ghysels et al.\ (2007), the penalized regression enables us to forecast 
the quarterly GDP using a large number of monthly predictors without imposing a distributed lag structure on the regression coefficients. 
We find that the forecasting performance of the penalized regression is better than 
that of the factor-based MIDAS (F-MIDAS) regression proposed by Marcellino and Schumacher (2010) and is competitive with the nowcasting model based on the state-space representation in real-time forecasting. 
The second application concentrates on  the oracle property. We investigate how well the penalized regression can screen a (hidden) fund manager's portfolio from large-dimensional NYSE stock price data. 
We construct artificial portfolios, and then we confirm the penalized regression using the SCAD-type penalty  effectively detects the relevant stocks 
that should be contained in the portfolio. These two convincing empirical applications motivate us to apply the penalized regression to macroeconomic time series broadly.


The remainder of the paper is organized as follows. Section \ref{sec2} specifies an ultrahigh-dimensional time series regression model and the estimation scheme. 
The statistical validity of the method is confirmed in Section \ref{sec3} by deriving the oracle inequality and the oracle property.  Section \ref{sec5} illustrates how 
we can apply the penalized regression for macroeconomic time series by two empirical analyses.
Section \ref{sec6} concludes. The proofs and miscellaneous results are collected in the Appendix.








\section{Regression Model} \label{sec2}

%




The regression model to be considered is 
\begin{align}\label{model2}
\v{y} = \v{X} \v{\beta}_0 + \v{u}, 
\end{align}
where $\v{y}=(y_1,\dots,y_T)^\top $ is a response vector, $\v{X}=(\v{x}_1,\dots,\v{x}_T)^\top$ is a covariate matrix with $\v{x}_t=(x_{t1},\dots, x_{tp})^\top$, $\v{u}=(u_{1},\dots,u_T)^\top$ is 
an error vector, and $\v{\beta}_0=(\v{\beta}_{0A}^\top ,\v{\beta}_{0B}^\top )^\top$ is a $p$-dimensional unknown sparse parameter vector with $\v{\beta}_{0A}=(\beta_{0,1},\dots,\beta_{0,s})^\top$ 
an $s$-dimensional vector of nonzero elements and $\v{\beta}_{0B}=\v{0}$. We also denote $j$th column vector of $ \v{X} $ by $\v{x}_j=(x_{1j},\dots, x_{Tj})^\top$. Further, we write $\v{X}=(\v{X}_A,\v{X}_B)$ 
corresponding to the decomposition of the parameter vector. Throughout the paper, we assume that for each $i$, $\{x_{ti}u_{t}\}_t$ is a martingale difference sequence with respect to an appropriate filtration.

The objective of the paper is how we construct an efficient $h$-step ahead forecast value of $y_{T+h}$ and how we select variables consistently when dimension $p$ is much larger than $T$. 
In such cases, $\v{X}$ may contain many irrelevant columns, so that the sparsity assumption on  $\v{\beta}_0$ may be appropriate. In this paper, we consider an {\it ultrahigh-dimensional} case, meaning that $p$ diverges 
sub-exponentially (non-polynomially). At the same time, the degree of sparsity $s$ may also diverge, but $s<T$ must be satisfied.
The estimation procedure should select a relevant model as well as consistently estimate the parameter vector. 
The estimator $\hat{\v{\beta}}$ is defined as a minimizer of the objective function
\begin{align}\label{obj}
Q_T(\v{\beta}) \equiv (2T)^{-1}\| \v{y}-\v{X}\v{\beta} \|_2^2 + \| p_\lambda(\v{\beta}) \|_1
\end{align}
over $\v{\beta} \in \mathbb{R}^p$, where $p_\lambda(\v{\beta})\equiv (p_\lambda(|\beta_1|),\dots,p_\lambda(|\beta_p|))^\top $ 
and $p_\lambda(v)$, for $v \geq 0$, is a penalty function indexed by 
a regularization parameter $\lambda (=\lambda_T) > 0$. 
The penalty function $p_{\lambda}$ takes forms such as the $\ell_1$-penalty (Lasso) by Tibshirani (1996), SCAD penalty by Fan and Li (2001), 
and MCP by Zhang (2010). These penalties belong to a family of so-called {\it folded-concave penalties} 
because of their functional forms. 
The statistical properties have been developed for models with a deterministic covariate and i.i.d.\ Gaussian errors in the literature on high-dimensional statistics. We thoroughly investigate these properties, while 
relaxing the assumptions sufficiently to include many time series models.

\begin{figure}
\begin{center}
\includegraphics[width=10cm,clip]{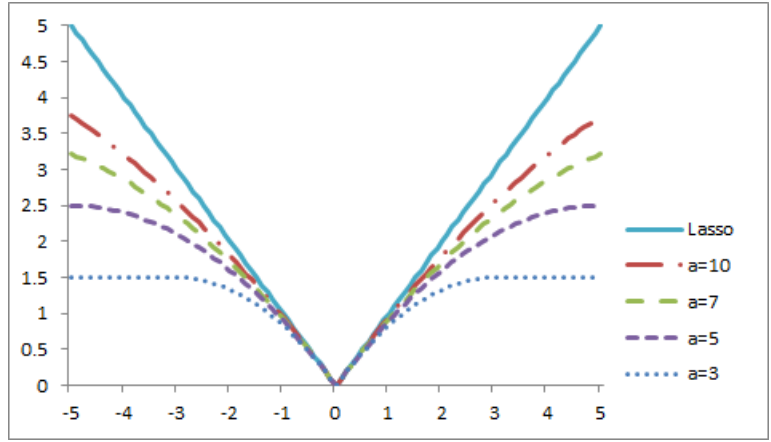}
\end{center}
\caption{Shape of Folded-Concave Penalties: MCP and Lasso.}
\label{fig:first}
\end{figure}

We introduce the three penalties to be used. Let $v$ denote a positive variable. 
The $\ell_1$-penalty is given by $p_\lambda (v)=\lambda v$, and we then obtain $p_\lambda' (v) = \lambda$
and $p_\lambda'' (v) = 0$.
The SCAD penalty is defined by 
\begin{align*}
p_\lambda (v) = 
\lambda v  1\{v \leq \lambda \} 
+\frac{a\lambda v - 0.5 (v^2 +\lambda^2)}{a-1}  1\{\lambda < v \leq a \lambda \} 
+\frac{\lambda^2(a^2-1)}{2(a-1)}  1\{ v >a\lambda \} .
\end{align*}
Its derivative is 
\begin{align*}
p_\lambda' (v) = \lambda \left\{ 1(v \leq \lambda) + \frac{(a\lambda -v)_+}{(a-1)\lambda} 1(v >\lambda) \right\},
\end{align*}
for some $a>2$. Then we have $p_\lambda'' (v)=-(a-1)^{-1}1\{ v \in (\lambda,a\lambda)\}$.
The MCP is defined by
\begin{align*}
p_\lambda (v) = 
\left(\lambda v -\frac{v^2}{2a}\right)1\{ v \leq a\lambda\} 
+\frac{1}{2}a\lambda^2 1\{ v > a \lambda \}.
\end{align*}
Its derivative is $p_\lambda' (v)= a^{-1}(a\lambda -v)_+$ for some $a \geq 1$.
Thus, we have $p_\lambda'' (v)=-a^{-1}1\{ v<a \lambda\}$. Figure \ref{fig:first} illustrates a shape of 
the MCP with several values of tuning parameters $a$ as well as that of  the Lasso.

\section{Two Theoretical Results}\label{sec3}

In this section, we establish two important theoretical results, the {\it oracle inequality} and {\it oracle property} for time series models. The oracle inequality gives optimal non-asymptotic error bounds for estimation and prediction 
in the sense that the error bounds are of the same order of magnitude up to a logarithmic factor as those we would have if we a priori knew the relevant variables (B\"ulmann and van de Geer, 2011). 
This result strongly supports the use of penalized regressions in terms of forecasting accuracy, even in ultrahigh-dimensional spaces. Note that we should remark that the inequality provides no information on model selection 
consistency; that is, it is not clear whether the penalized regression correctly distinguishes the relevant variables contained in the true model from 
the irrelevant ones. This issue is then addressed by the oracle property, which, in turn, states that the estimator exhibits model selection consistency. 
The existing results have shown the oracle inequality and the oracle properly under i.i.d.\ Gaussian errors and deterministic covariates, but in the paper we extend these results to apply to  time series models.



\begin{ass}\label{assporder}\rm 
We have $\log p = O(T^\delta)$ and $s=O(T^{\delta_0})$ for some constants $\delta,\delta_0 \in (0,1)$. 
\end{ass}

\begin{ass}\label{asspenalty}\rm 
Penalty function $p_\lambda(\cdot)$ is characterized as follows: 
\begin{itemize}
\item[(a)] $p_\lambda(v)$ is concave in $v\in [0,\infty)$ with $p_\lambda(0)=0$, 
\item[(b)] $p_\lambda(v)$ is nondecreasing, but $v \mapsto p_\lambda(v)/v$ ($v\not=0$) 
is nonincreasing  in $v\in [0,\infty)$,  
\item[(c)] $p_\lambda(v)$ has a continuous derivative $p_\lambda'(v)$ for $v\in (0,\infty)$ with $p_\lambda'(0+)= \lambda$, 
\item[(d)] There exists $\mu>0$ such that $p_\lambda(v)+\mu v^2/2$ is convex in $v\in [0,\infty)$. 
\end{itemize}
\end{ass}

\noindent
Assumption \ref{assporder} means that the dimensionality of the model, $p$, diverges sub-exponentially as $T$ goes to infinity. 
Assumption \ref{asspenalty} determines a family of folded-concave penalties that bridges $\ell_0$- and $\ell_1$-penalties.  The SCAD and MCP are included in this family. 
The $\ell_1$-penalty also satisfies this as the boundary of this class. 
It is known from Lemmas 6 and 7 of Loh and Wainwright (2014) that (d) is true provided that $\mu \geq 1/(a-1)$ for the SCAD and $\mu \geq 1/a$ for the MCP.

We define the gradient vector and Hessian matrix of $(2T)^{-1}\| \v{y}-\v{X}\v{\beta} \|_2^2$ as 
$G_T(\v{\beta})\equiv -\v{X}^\top (\v{y}-\v{X} \v{\beta})/T$ and $\v{H}_T\equiv \v{X}^\top \v{X}/T$, respectively. 
Denoting $\v{G}_{0T}\equiv G_T(\v{\beta}_0)$, we may write 
\begin{align*}
\v{G}_{0T}=-\frac{1}{T}
\begin{pmatrix}
\v{X}_A^\top  \v{u} \\
\v{X}_B^\top  \v{u} 
\end{pmatrix}
\equiv 
\begin{pmatrix}
\v{G}_{0AT} \\
\v{G}_{0BT} 
\end{pmatrix},
\quad \v{H}_T= \frac{1}{T}
\begin{pmatrix}
\v{X}_{A}^\top \v{X}_{A} & \v{X}_{A}^\top \v{X}_{B} \\
\v{X}_{B}^\top \v{X}_{A} & \v{X}_{B}^\top \v{X}_{B}
\end{pmatrix}
\equiv 
\begin{pmatrix}
\v{H}_{AAT} & \v{H}_{ABT} \\
\v{H}_{BAT} & \v{H}_{BBT}
\end{pmatrix}.
\end{align*}

\subsection{Oracle inequality}\label{subsecoracleineq}

We derive optimal non-asymptotic error bounds for estimation and prediction called the oracle inequality. In the literature, B\"ulmann and van de Geer (2011, Ch.\ 6) presented a complete guide for the inequality using the $\ell_1$-penalty 
with fixed predictors and i.i.d.\ normal errors. We extend the result in two ways. First, the inequality holds for the general model (\ref{model2}). Second, we prove the asymptotic equivalence of $\ell_1$- and the other folded-concave penalties 
characterized by Assumption \ref{asspenalty} in the sense that they satisfy the same rate. This indicates that the forecasting performance is asymptotically equivalent, irrespective of the folded-concave penalty used. 
We first derive the bounds under two high-level assumptions in Section \ref{oracleineq:general}. 
We next consider the conditions under which the two high-level assumptions are actually verified in a reasonable time series setting in Section \ref{oracleineq:special}. Related studies are introduced in Appendix \ref{subsecrelation}. 

\subsubsection{General result}\label{oracleineq:general}

We start with general but high-level assumptions: 

\begin{ass}\label{assevent}\rm 
There are a sequence $\lambda =o(1)$ and a positive constant $c_1$ such that $\mathcal{E}_1^c$, the complement of the event $\mathcal{E}_1 = \left\{ \| \v{G}_{0T} \|_\infty \leq \lambda/2 \right\}$, satisfies $P(\mathcal{E}_1^c) = O(p^{-c_1})$.
\end{ass}

\begin{ass}\label{assRSC}\rm 
There are a diverging sequence $m=o(T)$ and positive constants $c_2$ and $\gamma>\mu/2$ such that $\mathcal{E}_2^c$, the complement of the event $\mathcal{E}_2 = \left\{ \min_{\v{v}\in \mathbb{R}^p,~\|\v{v}\|_0\leq m} T^{-1}\|\v{X} \v{v}\|_2^2 / \|\v{v}\|_2^2 \geq \gamma \right\}$, satisfies $P(\mathcal{E}_2^c) = O(\exp(-c_2T))$. 
\end{ass}

\noindent
Assumption \ref{assevent} requires that the gradient vector $\v{G}_{0T}$ to behave less fluctuate and converge to zero with an appropriate rate determined by $\lambda$. For example, we should set $\lambda=O((\log p/T)^{1/2})$ for the case when $\v{u}$ is i.i.d.\ normal and $\v{X}$ is deterministic. 
Assumption \ref{assRSC} is a stochastic version of the {\it restricted strong convexity} studied by  Negahban et al.\ (2012). This prevents the minimum eigenvalue of the sub-matrix of Hessian matrix $\v{H}_{T}$ from being too small. 
These two assumptions fully control the randomness of the regression model, meaning that irrespective of the  dependence structure the model possesses, Theorem \ref{oracleineq} below holds as long as they are satisfied. 
The problem is what reasonable conditions on $\v{X}$ and $\v{u}$ satisfy Assumption \ref{assevent} and \ref{assRSC}. In fact, these can easily be verified for i.i.d.\ Gaussian $\v{u}$ and deterministic $\v{X}$. 
However, we may anticipate that it becomes quite unclear whether these assumptions hold or not once the model departs from such simple settings. 

Under the assumptions listed above, we  can derive the following result: 
\begin{thm} \label{oracleineq}
Let Assumptions  \ref{assporder}--\ref{assRSC} hold. 
Then, there exists a local minimizer $\hat{\v{\beta}}$ of $Q_T(\v{\beta})$ on $\{ \v{\beta} \in \mathbb{R}^p: \|\v{\beta}\|_0\leq m-s \}$ such that, with probability at least $1 - O(p^{-c_1}) - O(\exp(-c_2T))$, the following hold:
\begin{itemize}
\item[(a)] (Estimation error in $\ell_2$-norm) 
$\displaystyle \| \hat{\v{\beta}}-\v{\beta}_0 \|_2 
\leq \frac{6 s^{1/2}\lambda}{2\gamma-\mu}$,
\item[(b)] (Estimation error in $\ell_1$-norm) 
$\displaystyle \| \hat{\v{\beta}}-\v{\beta}_0 \|_1 
\leq \frac{24 s\lambda}{2\gamma-\mu}$,
\item[(c)] (Prediction loss) 
$\displaystyle T^{-1/2} \| \v{X}(\hat{\v{\beta}}-\v{\beta}_0) \|_2 
\leq \frac{9 s^{1/2}\lambda}{(2\gamma-\mu)^{1/2}}$.
\end{itemize}
\end{thm}

\noindent
If $2\gamma-\mu$ is assumed to be fixed, the error bounds converge to zero as long as $\lambda$ goes to zero relatively faster than $s^{1/2}$ or $s$. In a simple setting with i.i.d.\ Gaussian $u_t$ and fixed $X_t$, it is known that $\lambda$ should be 
given by $O((\log p/T)^{1/2})$ as mentioned before, leading to the explicit convergence rates $O((s\log p/T)^{1/2})$. This goes to zero provided that $\delta+\delta_0<1$. We observe later that the rates become slightly slower in a time series setting. 
%
%
Result (c) exhibits an optimal bound for the prediction loss in the $\ell_2$-norm in the sense of Bickel et al.\ (2009). 
This result justifies using any penalty function 
specified by Assumption \ref{asspenalty} when the aim is forecasting in the ultrahigh dimension. 
To understand the result, we consider a simplification in model (\ref{model2}) such that $\v{X}$ is deterministic, $\v{u}$ is i.i.d.\ with 
a unit variance, and $s=p<T$. Then, the squared risk of the OLS estimator $\hat{\v{\beta}}_{OLS}=(\v{X}^\top\v{X})^{-1}\v{X}^\top \v{y}$ becomes 
\begin{align*}
T^{-1}\E \| \v{X}(\hat{\v{\beta}}_{OLS}-\v{\beta}_0) \|_2^2 = T^{-1}\E[ \v{u}^\top \v{X} (\v{X}^\top \v{X})^{-1} \v{X}^\top \v{u}] = T^{-1}\tr \v{I} =s/T.
\end{align*}
Consider the case $p\geq T>s$. If we knew the true model $A$, we could choose the correct $s$ variables from $\v{X}$, leading to the risk $s/T$. However, since $A$ is unknown, the additional logarithm factor, which is regarded as the price to pay for 
not knowing $A$, is inserted. 

\subsubsection{When does the general result hold?}\label{oracleineq:special}

Theorem \ref{oracleineq} has established the non-asymptotic error bounds for the penalized regression estimators and prediction error under general, yet high-level, assumptions. Specifically, Assumptions \ref{assevent} and \ref{assRSC} should 
be verified for each model we attempt to employ. Here we consider a specific time series model. To consider a specific dependent model, we first strengthen the assumption on dimensionality:

\begin{ass}\label{assdim}\rm 
	The dimensionality is given by $\log p= \phi T^\delta$ and $s= \phi_0 T^{\delta_0}$ for some positive constants $\phi$, $\phi_0$, $\delta$, and $\delta_0$ such that $\delta + \delta_0 <1$. 
\end{ass}

In order to specify the processes of $ \v{X} $ and $ \v{u} $, we further assume in the same manner as Ahn and Horenstein (2013) that the covariate $\v{X}$ and the error $\v{u}$ are given by 
\begin{align}
\v{X} = \v{R}_X^{1/2} \v{Z}_X \v{\Sigma}_X^{1/2},~~~~~ \v{u} = \sigma_u\v{R}_u^{1/2} \v{z}_u,  \label{Xform}
\end{align}
where the random matrix $\v{Z}_X\in\mathbb{R}^{T\times p}$, random vector $\v{z}_u\in\mathbb{R}^{T}$, and deterministic matrices $\v{R}_X\in\mathbb{R}^{T\times T}$, $\v{R}_u\in\mathbb{R}^{T\times T}$, and $\v{\Sigma}_X\in\mathbb{R}^{p\times p}$ are characterized by the following assumption:

\begin{ass}\label{covariate1}\rm The following conditions hold: 
\begin{itemize}
\item[(a)] The entries of $\v{Z}_X$ and $\v{z}_u$ are i.i.d.\ standard normal random variables. 
\item[(b)] $\v{R}_X$, $\v{R}_u$, and $\v{\Sigma}_X$ are symmetric and positive definite non-random matrices, 
the minimum eigenvalues of which are bounded from below by positive constants $c_{R_X}$, $c_{R_u}$, and $c_\Sigma$, respectively. In addition, we set $c_R=c_{R_X} \wedge c_{R_u}$ and $\sigma_u>0$. 
\item[(c)]  $\v{R}_X^{1/2}\equiv\left(r_{st}^{(X)}\right)$ and $\v{R}_u^{1/2}\equiv\left(r_{st}^{(u)}\right)$ are lower triangular matrices whose elements satisfy $r_{tt}^{(X)} =r_{tt}^{(u)} = 1$ and $R_{sT}^{(X)} \vee R_{sT}^{(u)} =O(1) $  for all $s$, where $R_{sT}^{(X)}=\sum_{t=1}^T (r_{st}^{(X)})^2$ and $R_{sT}^{(u)}=\sum_{t=1}^T (r_{st}^{(u)})^2$. $ \v{\Sigma}_X^{1/2} \equiv\left(\sigma_{ij}^{(X)}\right)$ is a positive definite matrix that satisfies $ \sigma_{ii}^{(X)} = 1 $ and $ \Sigma_{pj}^{(X)} < \infty $ for all $j$, where $\Sigma_{pj}^{(X)}=\sum_{i=1}^p (\sigma_{ij}^{(X)})^2$. 
\end{itemize}
\end{ass}

Gaussianity in condition (a) can be weakened to sub-Gaussianity. 
Matrices in condition (c) are defined based on the Cholesky decomposition and Spectral decomposition under condition (b). Model (\ref{Xform}) with Assumption \ref{covariate1} covers a wide range of time series processes with cross-sectional dependences. A simple example of $ \v{R}^{1/2}_X $ and $ \v{\Sigma}^{1/2}_X $ is given by setting $ r^{(X)}_{t,t-1} = \theta_r $ and $ \sigma^{(X)}_{i,i-1} = \varphi_\sigma $ for some constants $\theta_r$ and $\varphi_\sigma$ satisfying $ | \theta_r | < 1 $ and $ |\varphi_\sigma | < \infty $ with other entries all zero. 
Obviously, this formulation satisfies condition (c) with reducing model \eqref{Xform} to an $\mathrm{MA}(1)$ process. 
Other weak stationary processes with cross-sectional dependences can be expressed in a similar manner.


\begin{prop}\label{propassevent}
Let Assumptions \ref{assdim} and \ref{covariate1} hold with 
$\lambda=c_0\log(pT) (\log p/ T)^{1/2}$,  with choosing positive constant $c_0$ such that $c_0\geq 16 c_{xu}$,  where $c_{xu} = \limsup_T \max_{t,i} \{R_{tT}^{(X)}\Sigma_{pi}, R_{tT}^{(u)}\sigma_u\} <\infty$. 
Then, Assumption \ref{assevent} is satisfied with $P(\mathcal{E}_1^c) \leq 6p^{-1}$. 
\end{prop}


\begin{prop}\label{propassRSC}
Let Assumptions \ref{assdim} and \ref{covariate1} hold with $m \leq \phi T^{1-\delta}$ and $\phi^2<1/2$. 
Then, Assumption \ref{assRSC} is satisfied with $\gamma = c_Gc_R/9$ and $P(\mathcal{E}_2^c) \leq 2\exp(-c_2T)$, where $c_2 = 1/2-\phi^2$. 
\end{prop}

Combining Propositions \ref{propassevent} and \ref{propassRSC} leads to the non-asymptotic bounds in the time series setting specified by Assumptions \ref{assdim} and \ref{covariate1}.

\begin{cor}\label{cor1}
Let Assumptions \ref{asspenalty}, \ref{assdim}, and \ref{covariate1} hold with the constants being the same as in Propositions \ref{propassevent} and \ref{propassRSC}. Then, there exists a local minimizer $\hat{\v{\beta}}$ of $Q_T(\v{\beta})$ on $\{ \v{\beta} \in \mathbb{R}^p: \|\v{\beta}\|_0\leq m-s \}$ such that, with probability at least $1 - 6p^{-1} - 2\exp\{-(1/2-\phi^2)T\}$, the the error bounds (a)--(c) of Theorem \ref{oracleineq} hold. 
\end{cor}

Corollary \ref{cor1} does not always imply the consistency. 
Once the condition $\delta+\delta_0 <1$ in Assumption \ref{assdim} is strengthened to $3\delta+\delta_0 <1$, the bounds of (a) and (c), given by $s^{1/2}\lambda=O\left(T^{\delta_0/2}\log(pT)(\log p/T)^{1/2}\right)$, converge to zero. Similarly, adding the condition $3\delta+2\delta_0 <1$ entails the bound of (b) converges to zero.

Compared to the conventional rate, $O\left((s\log p/T)^{1/2}\right)$, obtained with i.i.d.\ normal errors and fixed covariates, a slightly slower rate $O \left( (\log p T)  (s\log p/T)^{1/2} \right)$ arises for our time series model. We can interpret the additional factor $\log (pT) $ as an extra cost of departure from the independent Gaussian world. 
To understand this, the point is the behavior of the process $\{x_{ti}u_t\}$ for each $i$. If $u_t$ is i.i.d.\ Gaussian and $x_{ti}$ is deterministic, $\{x_{ti}u_t\}$ becomes a sequence of independent normal random variables. Hence, it is easy to control the tail probability $P(\|\v{G}_{0T}\|_\infty>\lambda)$ to be very small by using the inequality $P(|Z|>x)\leq \exp(-x^2/2)$ for $Z$ from $N(0,1)$ and for any $x>0$. 
Contrary to this conventional setting, ours assumes $x_t$ is stochastic, so that $\{x_{ti}u_t\}$ is no more independent Gaussian process. To evaluate the tail probability, we may use Azuma-Hoeffding's inequality together with the assumption that $\{x_{ti}u_t\}$ is a martingale difference sequence. In this case, we have to control the boundedness of $\{x_{ti}u_t\}$ at the same time, resulting in the additional factor $\log(pT)$ described above.


\subsection{Oracle property}\label{subsecoracle}

It is well known that the capacity of the Lasso for model selection is quite limited (e.g., Fan and Lv 2011). If we employ a SCAD-type penalty, however, a stronger and more desirable result on variable selection can be obtained. This result is called the oracle property, as studied first by Fan and Li (2001). 
The property admits $\hat{\v{\beta}}_A$ to be asymptotically equivalent to the maximum likelihood estimate obtained under the correct restriction $\v{\beta}_B=\v{0}$. To derive it under a time series setting, we need a different set of conditions; see Appendix \ref{assoracleprop}. Define $d (=d_T) \equiv \min_{j \in A}|\beta_{0,j}|/2$, 
$\v{I}_{0AA} \equiv T\E[ \v{G}_{0AT}\v{G}_{0AT}^\top ]$,   
and $\v{J}_{0AA} \equiv \E[ \v{H}_{AAT}]$.

Under assumptions in Appendix \ref{assoracleprop}, we will derive model selection consistency and appropriate rate of convergence. The role of Assumption \ref{assevent2} is essentially the  same as that of Assumption \ref{assevent}. The first condition in Assumption \ref{asspen} is a variant of 
the {\it beta-min condition} in B\"ulmann and van de Geer (2011, Ch.\ 7). This is necessary to distinguish the nonzero coefficient of relevant variables from zero though it seems stringent in the case of econometric modeling. The second condition $p_\lambda'(d)=0$ is key to achieve 
the oracle property. This is strong enough to exclude the $\ell_1$-penalty from Assumption \ref{asspenalty}. In fact, for the $\ell_1$-penalty, $p_\lambda'(v)=\lambda(>0)$ holds identically for all $v>0$. On the other hand, for the SCAD and MCP, this holds for a sufficiently 
large $T$ as long as $d/\lambda \rightarrow \infty$ is satisfied. Assumptions \ref{asslik0}--\ref{asslik} seem quite natural and are frequently used in stationary time series analysis. 
Assumption \ref{asslik_submat} restricts the asymptotic behavior of the lower-left $(p-s) \times s$ submatrix of $\v{H}_T$. This is essentially the same as condition (27) of Fan and Lv (2011).

Letting $\v{b} \in \mathbb{R}^s$ be such that $\| \v{b}\|_2^2=1$, 
we set $\xi_t\equiv \v{b}^\top \v{I}_{0AA}^{-1/2} \v{x}_{At}u_t$ and $\xi_{Tt}\equiv T^{-1/2}\xi_t$.  These can easily be shown to be a martingale difference sequence and martingale difference array, respectively. 
Note that $\sum_{t=1}^T\xi_{Tt}$ can also be written as $T^{1/2}\v{b}^\top \v{I}_{0AA}^{-1/2} \v{G}_{0AT}$. Assumption \ref{asslikasyn} is required to obtain the asymptotic normality. 
From Davidson (1994, Ch.\ 24), this leads to a central limit theorem of a martingale difference sequence. 
If $\xi_t$ is ergodic stationary, this is redundant (Billingsley, 1961).

\begin{thm}[oracle property]\label{exist}
Let Assumptions \ref{assporder}, \ref{asspenalty}, and \ref{assevent2}--\ref{asslik_submat} hold. 
Then, there exists a local minimizer $\hat{\v{\beta}}=(\hat{\v{\beta}}_A^\top ,\hat{\v{\beta}}_{B}^\top )^\top $
of $Q_T(\v{\beta})$ such that
\begin{itemize}
\item[(a)] (Sparsity) $\hat{\v{\beta}}_{B}=\v{0}$ with probability approaching one;
\item[(b)] (Rate of convergence) $\| \hat{\v{\beta}}_{A}-\v{\beta}_{0A} \|_2 =O_p((s/T)^{1/2})$.
\end{itemize}
In addition, if Assumption \ref{asslikasyn} holds, then 
for any $\v{b}\in\mathbb{R}^q$ satisfying $\|\v{b}\|_2^2=1$, we have
\begin{itemize}
\item[(c)] (Asymptotic normality)
$T^{1/2}\v{b}^\top  \v{I}_{0AA}^{-1/2} \v{H}_{AAT}^\top (\hat{\v{\beta}}_A-\v{\beta}_{0A}) \rightarrow _d N( 0,1)$.
\end{itemize}
\end{thm}

The oracle property means that the model selection is consistent in the sense of (a) and (b). 
Moreover, as is understood by result (c), the estimator has the same asymptotic efficiency 
as the (infeasible) MLE obtained with advance knowledge of the true submodel. Based on these results, we can estimate 
ultrahigh-dimensional models without irksome tests for zero restrictions on the parameters or 
an exhaustive search using information criteria.



\section{Empirical Examples}\label{sec5}

According to the theoretical results given in the previous sections, the penalized regression can have two desirable properties: the oracle inequality and the oracle property. In this section, we provide 
two empirical examples that motivate how well the penalized regression works in macroeconometric analyses. The first forecasts the quarterly real U.S.\ GDP 
with a large number of monthly macroeconomic predictors, and the second screens portfolio from a large number of potential securities using NYSE stock price data.

\subsection{Forecasting quarterly U.S. GDP with a large number of predictors } \label{penalized regression MIDAS}

\subsubsection{Penalized MIDAS regression model} 

In this section, we illustrate how to apply the penalized regression model to macroeconomic  time series using the MIDAS forecasting regression.  
The MIDAS regression model  was originally proposed by Ghysels et al.\ (2007) and is now one of standard tool for forecasting with mixed-frequency
data, as well as the now-casting model based on the state-space representation (e.g., Giannone et al., 2008; Ba\'{n}bura and Modugno, 2013). 
The original (or basic) MIDAS regression model has an advantage of describing a forecasting regression model in a simple and parsimonious way of a distributed lag structure with a few hyperparameters. 
However, the original MIDAS regression model would not be suitable for a situation where the number of predictors in the model is very large. For example, consider the original MIDAS regression model 
with $K$ hyperparameters and $N$ macroeconomic time series. Then, the total number of parameters in the original MIDAS regression model remains $ NK + 1 = O(N) $. Thus, it invokes a serious efficiency loss if $ N $ is large or even it makes the model inestimable. 
On the other hand, the penalized regression enables us to  estimate the MIDAS regression model without imposing the distributed lag structure on the regression coefficients. 
Moreover, Theorem 1 implies that the forecast value obtained by the penalized regression is reliable. 
In the following, we link the penalized regression model (\ref{model2}) to the MIDAS regression model without parameter restrictions, and consider to forecast quarterly U.S.\ GDP with the monthly macroeconomic 
data using the penalized regression. 

Let $\{ y_t,\v{x}_{t/m}^{(m)}\}$ be the MIDAS process in line with  Andreou et al.\ (2010), where the 
scalar $y_t$ is the low-frequency variable observed at $t=1,\dots,T$, 
and the $N$-dimensional vector $\v{x}_{t/m}^{(m)}=\left( 1,x_{2,t/m}^{(m)},\dots, x_{N,t/m}^{(m)} \right)^\top $ is a set of higher-frequency variables observed  $m$ times between $t$ and $t-1$. 
For example, $ m = 3 $ if we forecast a quarterly variable with monthly predictors.
We consider the  $ h $-step-ahead mixed-frequency forecasting regression model with $\ell $ lags, 
\begin{align}\label{MIDASmodel}
y_{t} = \v{x}_{t-h}^{\top} \, \v{\beta}_0 + u_t, \qquad t=1,\dots,T,  
\end{align}
where $\v{x}_{t-h} =\left( 1, \v{x}_{2,t-h,\ell}^{(m)}, \dots, \v{x}_{N,t-h,\ell}^{(m)} \right)^\top $ with $ \v{x}_{k,t-h,\ell}^{(m)} = \left(x_{k,t-h/m}^{(m)}, x_{k,t-h-1/m}^{(m)} \dots, x_{k,t-h-\ell/m}^{(m)}  \right)^\top  $ for $ k=2,3,\dots,N $, 
 $\v{\beta}_0=(\beta_{0,1},\dots,\beta_{0,N\ell+N-\ell})^\top $ is the parameter vector and $u_t$ is an error term. Here the case $ h < 1$~($ h =0, 1/m, 2/m, \dots, (m-1)/m $) corresponds to nowcast; we forecast a low-frequency variable with the ``latest" high-frequency variables released between $t-1$ and $t$.
For instance, if we consider a quarterly/monthly ($m=3$) case, $ h = 0~(1/3) $ means that we forecast a quarterly variable in 2015Q2 with monthly data in June (May) 2015 or later. 
Note that model (\ref{MIDASmodel}) has the same structure as (\ref{model2}) with $ p := (N-1)(\ell+1)+1 = N\ell + N-\ell $ but it differs from the original MIDAS regression model by Ghysels et al.\ (2007); our model does not employ the distributed lag structure on $ \v{x}_{t-h} $ while they used  $\v{x}_{t-h} (\theta)=\left( 1, x_{2,t-h}^{(m)}(\theta_2), \dots, x_{N,t-h}^{(m)}(\theta_N) \right)^\top $ instead of $ \v{x}_{t-h} $ such that  
$ x_{k,t}^{(m)} (\theta_k) = \sum_{j=1}^\ell w_{j,k}(\theta_k) L^{j/m} x_{k,t/m}^{(m)} $ 
for $ k=2,\dots,p, $ where $w_{j,k}(\theta_k) \in (0,1)$ and $\sum_{j=1}^\ell w_{j,k}(\theta_k)=1$. As mentioned above,  the original  MIDAS model crucially depends on the restrictive distributed lag structure and cannot reduce the total number of the 
parameters to be estimated effectively if $N$ is very large.  Alternatively, the MIDAS regression that minimizes the penalized loss 
can estimate $ \v{\beta}_0 $ and forecast $y_t$ without the distributed lag structure.

In a macroeconomic forecasting point of view, it is natural to consider that there is a small set of key predictors that contain rich information to forecast $y$ while there are lots of redundant predictors. To reduce accumulation of estimation errors, we should model $y$ only by using the key predictors. Although the redundant predictors would have ``non-zero" forecasting power, the penalized regression makes their coefficient estimates zero as an approximation. In other words, we can say that the sparsity assumption claims there exist ``targeted predictors" for $y$ (Bai and Ng, 2008).

         Hereafter, we call the MIDAS regression model estimated by the penalized regression  as ``penalized MIDAS regression." 
We also note that as a method related to our penalized MIDAS regression, Marsilli (2014) proposes a MIDAS regression model with a penalized regression. 
However, he employed the original  MIDAS parsimonious parameterization, which completely differs from our model in terms of parameterization as we stressed above.

\subsubsection{Data} 

U.S.\ quarterly real GDP growth is taken from the FRED database. 
The sample period is from 
1959Q4 to 2016Q2. We retrieve 117 U.S.\ monthly macroeconomic time series ($N=117$) from the FRED--MD database and the series are appropriately detrended according to a guideline given in McCracken 
and Ng (2015). Note that the FRED--MD database originally contains a total of 128 series, but we remove 11 series due to the following reasons: the CBOE S\&P 100 Volatility Index (VXOCLSx), 
Consumer sentiment index (UMCSENTx), Trade weighted U.S.\ dollar index of  major currencies (TWEXMMTH),  
New orders for nondefense capital goods (ANDENOx), New orders for consumer goods (ACOGNO), and New private housing permits (PERMIT, PERMITNE, PERMITMW, PERMITS, PERMITW) have no observations 
from 1959. Furthermore, our preliminary inspection found that Reserves of depository institutions nonborrowed (NONBORRES) contained extreme changes in February 2008, which would contaminate our analysis. 
The sample period of the detrended monthly series is from March 1959 (1959:3) to June 2016 (2016:6).

%


\subsubsection{Forecasting Strategy} 

We evaluate the out-of-sample forecasting performance by mean squared forecast errors (MSFE) in the evaluation period from 2000Q1 to 2016Q2. 
The parameter estimates are obtained from each estimation period; the initial period is 1959Q4--1999Q4 and the next one extends the end point to 2000Q1 
with the starting point 1959Q4 being fixed. For example, the initial forecast error in 2000Q1 is calculated using the estimates from the initial estimation period 
1959Q4--1999Q4, and the second forecast error in 2000Q2 uses the estimates from the second estimation period 1959Q4--2000Q1. 
We suppose that the forecast regression consists of eight lags ($\ell =8 $), so that the total number of parameters for the forecasting regression to be estimated is $ N\ell + N-\ell = 117 \times 8 + 117 - 8 = 1045$, including 
a constant term.  The penalized  MIDAS regression is expected to be robust to a choice of  $ \ell $,  as long as we choose $ \ell $ to be moderately large, because the penalized regression conducts model selection as well as parameter estimation.     
To investigate the forecasting performance of the penalized  MIDAS regression model with a variety of horizons, we examine cases where 
$ h=0, 1/3, 2/3, 1, 4/3, 5/3, 2 $ in the same manner as Clements and Galv\~{a}o (2008) and Marcellino and Schumacher (2010). 
The cases $ h=0, 1/3, $ and $ 2/3 $ correspond to nowcasting 
in the sense that we forecast contemporaneous or very short-forecast-horizon quarterly GDP growth using monthly series before the official announcement of the GDP, 
while the case $ h= 2 $ is a forecast with a relatively long horizon. The sample size of the estimation period $T$ gradually increases and varies 
depending on $h$; for example, $T$ ranges from 161 to 227 if $ h =0 $, and from 159  to 225 if $ h = 2 $.    

Finally, we need to determine the values of the tuning parameters, $a$ and $\lambda$, in advance of the penalized  MIDAS regression regression. 
Following the guidelines by Breheny and Huang (2011, pp.\ 19 and 21) with our preliminary inspection of the overall samples, we set $ a = 12 $ for the SCAD and MCP, 
although the performance could be improved by a more careful choice. 
The value of $\lambda$ is selected by 10-fold cross-validation. The validity was confirmed by  Uematsu and Tanaka (2015). 
All estimations for the penalized regression are conducted using
 R 3.2.1 with the {\tt ncvreg} package of Breheny and Huang (2011).   



\subsubsection{Forecast performance} \label{forecastperformance}

To measure the performance appropriately, we consider two types of datasets. 
The first is a complete dataset, that is, there are no missing values in the dataset. The second is a real-time dataset, which has jagged/ragged edge 
pattern due to the publication lag of the series. 
 
\subsubsection{Forecast performance in complete data} \label{forecastperformance1}

We use data from 1959Q4--2016Q1 for the GDP and 1959:3 to 2016:3 to retrieve a complete dataset. We consider 
the following three evaluation periods: ($i$) Overall (2000Q1--2016Q1), ($ii$) 1st subsample (2000Q1--2007Q4), and ($iii$) 2nd subsample (2008Q1--2016Q1). 
This is because the unprecedented turmoil of the U.S.\ economy stemming from the subprime mortgage crisis and the ensuing collapse of Lehman Brothers in 2008
 would introduce parameter instability that would distort the forecast evaluation. 
As a result, we consider the forecast performance of the penalized regression in complete data from a total of $ 65 $ (overall), $ 32 $ (1st subsample) and 
$ 33 $ (2nd subsample) squared forecast errors, respectively.

Tables \ref{MSEall}--\ref{MSE2} report the mean squared forecast errors (MSFE) of the penalized  MIDAS regression with the SCAD, MCP, and Lasso, and their two competitors 
in the overall sample, 1st subsample, and 2nd subsample, respectively. In the tables, the median squared forecast errors are also shown in parentheses to remove contamination  
by outliers. The all values are relative values compared to a naive AR(4) forecast. The two competitors 
are the factor MIDAS (denoted ``Factor" in the tables) proposed by Marcellino and Schumacher (2010) and  the two-step penalized regression (post-OLS) procedure (denoted as 
``post-MCP," ``post-SCAD," ``post-Lasso" in the tables) proposed by Belloni and Chernozhukov (2013). The factor MIDAS is expected to be one of the strong competitors since the factor-based forecast is found to perform well in forecasting real variables (e.g., Stock and Watson, 2002, 2012; De Mol et al., 2008.). 
The factor MIDAS considered here is based on the basic MIDAS structure with  
the exponential Almon lag structure of two hyperparameters. The number of factors is assumed to be seven ($r=7$) based on the information criterion $ IC_{p2}$ by Bai and Ng
 (2002). Although we can consider the {\it unrestricted} Factor MIDAS as in Marcellino and Schumacher (2010), which is free from the 
distributed lag structure, we do not employ it  because of its intractability caused by high dimensionality. 
The two-step procedure using the Lasso is known as the {\it OLS post-Lasso}. Belloni and Chernozhukov (2013) showed it could perform at least as well as the Lasso and could be better in some cases. 
We also consider the two-step procedure using the MCP and SCAD penalties. 

First, we consider the nowcasting ($ 0 \leq h < 1 $) cases. Table \ref{MSEall} shows that all methods are much better than the naive AR(4) forecast, but that the penalized  MIDAS regression outperforms 
the factor MIDAS and the two-step procedures in the overall sample with a few exceptions, in terms of both the mean and median squared forecast errors. The two-step procedures work well in terms 
of MSFE, but  do not seem to work well in the median measure since they are frequently beaten by the naive AR(4) forecast. We also find that the MSFE of the factor when $ h = 1/3 $ is much 
worse than other methods, owing to  outliers of forecast values around the subprime mortgage crisis. 
Tables \ref{MSE1} and \ref{MSE2} show the forecasting performance for the first and second subsamples, respectively. In first subsample, the penalized  MIDAS regression does not necessarily work well; it performs well when $ h = 0 $, but
worse than the factor MIDAS when $ h=1/3 $ and $ 2/3 $. However, we also find that the penalized  MIDAS regression performs well and completely dominates the factor MIDAS and the two-step procedures in the second subsample 
in terms of both mean and median measures. Thus, it can be said that the penalized  MIDAS regression is more robust than the other methods in terms of structural instability. 
Furthermore, we find that the MSEs of the two-step procedure are worse than those of the penalized  MIDAS regression, overall. Thus, the two-step procedure does not provide effective efficiency gains in our situation. A probable reason is that the total number of regressors 
in the second-step OLS regression does not become effectively small when we assume a long-length lag structure in the model even if variable ``screening" 
is conducted in the first step. This would make the efficiency losses arising from estimating many parameters more serious than estimating penalized  MIDAS regression directly.   
Next, we turn to the forecast performance when $ h \geq 1 $. The tables show that all the methods have similar forecast performances; they perform well when $h = 1$, 
however, when $ h > 1 $, they  are all beaten by  the AR(4) forecast. The results are not surprising because 
 Clements and Galv\~{a}o (2008) and  Marcellino and Schumacher (2010) also find the same results.  Hence, our results show that the penalized MIDAS has a good forecast performance in a very short horizon, especially in the presence of instability, 
although it is not necessarily a primary tool for a forecast with a relatively long horizon. However, we can conclude that penalized  MIDAS regression is an effective tool for   
forecasting with mixed-frequency data because our main interest in forecasting with mixed-frequency data is nowcasting where low-frequency data are not available.


\begin{table}[t!]
\renewcommand{\arraystretch}{0.8}
\caption{Mean/Median Forecast Errors of the forecasts in complete data [Overall Sample]}
\label{MSEall}
\begin{center}
\begin{tabular}{rrrrrrrr} 
 & $ h=0 $  & $ h=1/3 $  & $ h=2/3 $  & $ h=1 $  & $ h=4/3 $  & $ h=5/3 $  & $ h=2 $  \\\hline
MCP & 0.58 & 0.50 & 0.53 & 0.80 & 1.17 & 1.35 & 1.34 \\
(median) & (0.80) & (0.79) & (0.80) & (0.57) & (1.32) & (1.16) & (1.18) \\
SCAD & 0.59 & 0.53 & 0.61 & 0.79 & 1.15 & 1.34 & 1.34 \\
(median) & (0.76) & (0.86) & (0.75) & (0.60) & (1.28) & (1.20) & (1.19) \\
Lasso & 0.56 & 0.56 & 0.60 & 0.79 & 1.15 & 1.33 & 1.34 \\
(median) & (0.80) & (0.89) & (0.70) & (0.61) & (1.27) & (1.17) & (1.30) \\
Factor & 0.83 & 2.12 & 0.89 & 0.75 & 1.89 & 1.25 & 1.25 \\
(median) & (1.11) & (0.89) & (0.81) & (1.00) & (1.12) & (1.62) & (1.73) \\
post-MCP & 0.79 & 0.62 & 0.61 & 0.82 & 1.16 & 1.43 & 1.50 \\
(median) & (1.48) & (1.19) & (1.01) & (0.79) & (1.09) & (1.63) & (2.06) \\
post-SCAD & 0.79 & 0.60 & 0.56 & 0.81 & 1.21 & 1.35 & 1.46 \\
(median) & (1.20) & (1.27) & (0.81) & (0.88) & (1.17) & (1.41) & (1.59) \\
post-Lasso & 0.82 & 0.61 & 0.93 & 0.81 & 1.17 & 1.36 & 1.46 \\
(median) & (1.35) & (1.14) & (0.92) & (0.88) & (1.02) & (1.48) & (1.55) \\
\end{tabular}
\end{center}
{\footnotesize Note) All values are relative values to AR(4) forecast. Values in parentheses are median forecast errors. } 
\end{table}

\begin{table}[t!]
\renewcommand{\arraystretch}{0.8}
\caption{Mean/Median Forecast Errors of the forecasts in complete data [1st Subsample]}
\label{MSE1}
\begin{center}
\begin{tabular}{rrrrrrrr} 
 & $ h=0 $  & $ h=1/3 $  & $ h=2/3 $  & $ h=1 $  & $ h=4/3 $  & $ h=5/3 $  & $ h=2 $  \\\hline
MCP & 0.76 & 0.74 & 0.70 & 0.93 & 1.03 & 1.29 & 0.76 \\
(median) & (0.84) & (0.94) & (1.15) & (0.90) & (1.37) & (1.77) & (0.84) \\
SCAD & 0.77 & 0.78 & 0.71 & 0.92 & 1.03 & 1.27 & 0.77 \\
(median) & (0.81) & (0.97) & (1.07) & (0.83) & (1.28) & (1.53) & (0.81) \\
Lasso & 0.74 & 0.77 & 0.76 & 0.92 & 1.03 & 1.27 & 0.74 \\
(median) & (0.59) & (0.88) & (1.15) & (0.84) & (1.28) & (1.51) & (0.59) \\
Factor & 0.86 & 0.69 & 0.60 & 0.86 & 1.06 & 1.76 & 0.86 \\
(median) & (0.98) & (0.86) & (0.86) & (1.27) & (1.46) & (3.34) & (0.98) \\
post-MCP & 0.95 & 1.09 & 0.92 & 1.09 & 1.23 & 1.60 & 0.95 \\
(median) & (1.52) & (1.31) & (1.52) & (1.28) & (1.51) & (2.41) & (1.52) \\
post-SCAD & 1.21 & 1.00 & 0.76 & 1.04 & 1.07 & 1.55 & 1.21 \\
(median) & (0.75) & (1.22) & (1.06) & (1.59) & (1.69) & (1.52) & (0.75) \\
post-Lasso & 1.33 & 1.03 & 1.51 & 1.07 & 1.06 & 1.60 & 1.33 \\
(median) & (1.19) & (1.40) & (1.54) & (1.60) & (1.35) & (1.52) & (1.19) \\
\end{tabular}
\end{center}
{\footnotesize Note) All values are relative values to AR(4) forecast. Values in parentheses are median forecast errors. } 
\end{table}

\begin{table}[t!]
\renewcommand{\arraystretch}{0.8}
\caption{Mean/Median Forecast Errors of the forecasts in complete data [2nd Subsample]}
\label{MSE2}
\begin{center}
\begin{tabular}{rrrrrrrr} 
 & $ h=0 $  & $ h=1/3 $  & $ h=2/3 $  & $ h=1 $  & $ h=4/3 $  & $ h=5/3 $  & $ h=2 $  \\\hline
MCP & 0.49 & 0.38 & 0.45 & 0.73 & 1.24 & 1.38 & 1.37 \\
(median) & (0.86) & (0.79) & (0.79) & (0.54) & (1.32) & (0.96) & (1.16) \\
SCAD & 0.50 & 0.41 & 0.57 & 0.73 & 1.21 & 1.37 & 1.38 \\
(median) & (0.76) & (0.89) & (0.66) & (0.56) & (1.35) & (1.20) & (1.30) \\
Lasso & 0.48 & 0.47 & 0.52 & 0.73 & 1.21 & 1.37 & 1.38 \\
(median) & (1.09) & (1.06) & (0.68) & (0.59) & (1.35) & (1.15) & (1.30) \\
Factor & 0.82 & 2.81 & 1.03 & 0.69 & 2.30 & 0.99 & 1.19 \\
(median) & (1.54) & (1.39) & (0.94) & (1.07) & (1.09) & (1.53) & (1.43) \\ 
post-MCP & 0.71 & 0.40 & 0.46 & 0.69 & 1.13 & 1.35 & 1.38 \\
(median) & (1.56) & (1.28) & (0.84) & (0.72) & (0.89) & (1.56) & (2.24) \\
post-SCAD & 0.59 & 0.40 & 0.47 & 0.70 & 1.28 & 1.25 & 1.36 \\
(median) & (1.78) & (1.48) & (0.82) & (0.72) & (1.14) & (1.60) & (1.54) \\
post-Lasso & 0.57 & 0.41 & 0.65 & 0.68 & 1.23 & 1.24 & 1.32 \\
(median) & (1.65) & (1.09) & (0.77) & (0.72) & (1.14) & (1.60) & (1.46) \\
\end{tabular}
\end{center}
{\footnotesize Note) All values are relative values to AR(4) forecast. Values in parentheses are median forecast errors. } 
\end{table}


\subsubsection{Forecast performance in real-time data} \label{forecastperformance2}

Section \ref{forecastperformance1} reveals that the penalized regression behaves well  in nowcasting with a complete data.  However, when we actually conduct real-time forecasting of quarterly GDP with monthly data, 
a complete dataset is not available because of possible publication lags of the series. Thus, we must face an incomplete dataset so called ``jagged (ragged)-edge" dataset,  that contains missing values in some latest months. Then we investigate 
how well the forecast with penalized regression works with the real-time data. It should be mentioned that in our experiment, strictly speaking,  we consider ``pseudo" real-time forecasting; we suppose each monthly data for all evaluation 
periods have the same  jagged (ragged)-edge pattern  as of the 2016-08 version of  the FRED-MD. For example,  Real manufacturing and trade industry sales (CMRMTSPLx) and the Help-wanted index (HWI) have one and four month  missing values owing 
to publication lags in the 2016-08 version, respectively. Then we suppose the data for all estimation periods have the same jagged-edge patterns even if our dataset contains complete data for those periods. Moreover, we assume no data revisions occur in our dataset.

Tables \ref{MSEalljag}--\ref{MSE2jag} show the relative MSFEs of the penalized regression and the state-space ML estimator proposed by Ba\'{n}bura and Modugno (2014) 
in the real-time overall sample (2000Q1--2016Q2),  1st subsample (2000Q1--2007Q4), and 2nd subsample (2008Q1--2016Q2), respectively. The tables omit the results for $ h > 1 $ and concentrate on the nowcast situation  ($ 0 \leq h \leq 1 $)  because 
the real-time forecasting is meaningful only in a very short horizon. The state-space ML estimation enables us to handle real-time mixed frequency data  by embedding missing patterns of data in the model; 
see Ba\'{n}bura and Modugno (2014) for details. On the other hand,  the penalized regression requires an interpolated dataset to obtain the forecast values. Thus, we employ an interpolation method based on the EM algorithm proposed by Stock and Watson (2002).

\begin{table}[t!]
\renewcommand{\arraystretch}{0.8}
\caption{Mean/Median Forecast Errors of the forecasts in jagged-edge data [Overall sample]}
\label{MSEalljag}
\begin{center}
\begin{tabular}{rrrrr} 
 & $ h=0 $  & $ h=1/3 $  & $ h=2/3 $  & $ h=1 $  \\\hline
MCP & 0.58  & 0.50  & 0.54  & 0.80  \\
(median) & (0.82) & (0.81) & (0.88) & (0.62) \\
SCAD & 0.60  & 0.53  & 0.62  & 0.80  \\
(median) & (0.79) & (0.88) & (0.81) & (0.67) \\
Lasso & 0.57  & 0.57  & 0.61  & 0.80  \\
(median) & (0.92) & (0.93) & (0.78) & (0.67) \\
State-Space ML & 0.47  & 0.49  & 0.66  & 0.84  \\
(median) & (0.70) & (0.71) & (1.00) & (1.02) \\
\end{tabular}
\end{center}
{\footnotesize Note) All values are relative values to AR(4) forecast. Values in parentheses are median forecast errors. } 
\end{table}

From the tables, we first find  the effects of the jagged-edge and interpolation on the forecast accuracy of the penalized regression are negligible since they  do not  essentially affect  the mean/median squared forecast errors values compared with 
the results in Tables \ref{MSEall}--\ref{MSE2}. Second, we see that the penalized regression performs well in the overall and 2nd subsample; it beats the state-space ML when $ h = 2/3 $ and $ 1 $ in both the mean/median measures, and performs 
as well as the state-space ML when $ h = 1/3 $ while it does not relatively work well in the 1st subsample as in the complete data case. The state-space ML is expected to have higher forecasting performance  than the penalized 
regression because the state-space ML is based on a system equation with richer information while the penalized regression relies on a single equation. 
However, this would not be true when a model misspecification is present, as Bai et al.\ (2013) claimed. Then, our results that reveal  the penalized regression can be compete with the state-space ML in terms of forecasting accuracy imply that the system equation 
contains a certain level of the misspecification. Moreover, it should be mentioned that the penalized regression is much simpler and rapid than the state-space ML in obtaining the forecast values. Since the dimension of the state-space model can be very large
 when we forecast with mixed frequency (117 dimensional state-space models with 40 latent factors in our case), the estimation is much computationally demanding and 
time consuming (roughly eight times longer than the penalized regression). Furthermore, the estimated values can be unstable if we consider to apply the state-space ML to a dataset with larger $N$ and/or $r$.                 

Although we do not examine them here, the Ridge regression and the Bayesian VAR (BVAR) would be potential alternatives to the state-space ML (e.g., De Mol et al., 2008; and Schorfheide and Song, 2015).
However, they are also computationally demanding (the BVAR requires more than 100,000 parameter estimation in our case) and their theoretical properties have not been investigated yet under ``ultra"high- dimensionality (i.e. $p$ diverges at a sub-exponential rate).    

\begin{table}[t!]
\renewcommand{\arraystretch}{0.8}
\caption{Mean/Median Forecast Errors of the forecasts in jagged-edge data [1st Subsample]}
\label{MSE1jag}
\begin{center}
\begin{tabular}{rrrrr} 
 & $ h=0 $  & $ h=1/3 $  & $ h=2/3 $  & $ h=1 $  \\\hline
MCP & 0.76  & 0.74  & 0.70  & 0.93  \\
(median) & (0.84) & (0.94) & (1.15) & (0.90) \\
SCAD & 0.77  & 0.78  & 0.71  & 0.92  \\
(median) & (0.81) & (0.97) & (1.07) & (0.83) \\
Lasso & 0.74  & 0.77  & 0.76  & 0.92  \\
(median) & (0.59) & (0.88) & (1.15) & (0.84) \\
State-Space ML & 0.62  & 0.67  & 0.71  & 0.94  \\
(median) & (0.72) & (0.72) & (0.65) & (1.17) \\
\end{tabular}
\end{center}
{\footnotesize Note) All values are relative values to AR(4) forecast. Values in parentheses are median forecast errors. } 
\end{table}

\subsection{Screening Effective Portfolio from a Large Number of Potential Securities}

Recent studies on portfolio selection have focused on the penalized regression because it plays a crucial role 
in constructing a portfolio when there are a large number of potential stocks. Brodie et al.\ (2009) find out 
the penalized regression is useful in selecting optimal portfolio in terms of the out-of-sample performance 
measured by the Sharpe ratio; Fan et al.\ (2012) introduced gross-exposure constraints to admit short sales in the estimation of 
an optimal portfolio; Carrasco and Noumon (2012) focused on estimating a precision matrix of returns. They found the penalized regression is quite useful to stabilize the 
estimation of the covariance matrix and provided better finite sample performances than traditional methods. 

To the best of our knowledge, the existing literature concerning applications of the penalized regression to portfolio selection focused on yieldability. 
However, it seems interesting to examine the consistent estimation of weights of the portfolio; that is, screening how fund managers construct 
their portfolio from a large number of securities is valuable. 
Unlike the other high-dimensional estimation methods, such as the factor and the Ridge, the SCAD-type penalized regression enables us to screen 
their portfolio from a large dataset of stock prices. In this section, we examine how well the penalized 
regression usefully works in this direction using a large NYSE stock price dataset.

\begin{table}[t!]
\renewcommand{\arraystretch}{0.8}
\caption{Mean/Median Forecast Errors of the forecasts in jagged-edge data [2nd Subsample]}
\label{MSE2jag}
\begin{center}
\begin{tabular}{rrrrr} 
 & $ h=0 $  & $ h=1/3 $  & $ h=2/3 $  & $ h=1 $  \\\hline
MCP & 0.50  & 0.39  & 0.46  & 0.74  \\
(median) & (1.06) & (0.82) & (0.81) & (0.55) \\
SCAD & 0.51  & 0.42  & 0.58  & 0.74  \\
(median) & (0.83) & (0.90) & (0.71) & (0.65) \\
Lasso & 0.49  & 0.48  & 0.53  & 0.74  \\
(median) & (1.13) & (1.10) & (0.69) & (0.65) \\
State-Space ML & 0.41  & 0.40  & 0.63  & 0.79  \\
(median) & (0.84) & (0.76) & (1.05) & (1.12) \\
\end{tabular}
\end{center}
{\footnotesize Note) All values are relative values to AR(4) forecast. Values in parentheses are median forecast errors. } 
\end{table}

\subsubsection{Construction of Portfolio} 

Suppose a fund manager faces $p$ potential stocks, where $ x_{it} $ is the rate of return of the $i$th ($i=1,2,\dots,p$) stock at time $t$.   
Let $ \v{x}_t = [x_{1t}, x_{2t}, \dots, x_{pt}]^{\top} $ be the $p$-dimensional rate of the return vector at $t$ and 
$ \v{\omega}_0 $ be the $p$-dimensional weight vector of the portfolio that satisfies  $ \|\v{\omega}_0 \|_0 = s~(\ll p) $, $ \v{\iota}' \v{\omega}_0 = 1 $ and $ \|\v{\omega}_0 \|_1 = \zeta_w $, 
where $ \zeta_w \in [1,\infty) $ and $ \v{\iota} $ is a $p$-dimensional vector with all elements being one. That is, the portfolio is constructed by $s$ stocks from $p$ potential stocks. 
We assume the fund manager constructs her portfolio as
\begin{align}
y_{t} = \v{x}_{t}^{\top} \, \v{\omega}_0 + u_t, \qquad t=1,\dots,T, \label{portfolio} 
\end{align} 
where $ u_t $ is a ``miscellaneous" component that includes all assets in the portfolio other than stocks, such as T-bills and corporate bonds. Further we 
assume that $ \v{x}_t $ and $ u_t $ are independent of each other and $ u_t \sim i.i.d. N(0,\sigma^2_u $), where $ \sigma^2_u = T^{-1} \v{\omega}^{\top}_{0A} \v{X}^{\top}_{A} \v{X}_{A} \v{\omega}_{0A} / \mathrm{SNR} $, 
$ \v{\omega}_{0A} $ is a nonzero $s$-dimensional subvector of $ \v{\omega}_0 $, $ \v{X}_A $ is $ T \times s $ submatrix of $ \v{X} $ that corresponds to $ \v{\omega}_{0A} $,  and $ \mathrm{SNR} = V(\v{x}_{t}^{\top} \, \v{\omega}_0)/ V(u_t) $. 
Although we might consider the case in which $ \v{x}_t $ and $ u_t $  are dependent by extending the results of Fan and Liao (2014), this is beyond the scope of our research, and we regard $ \v{x}_t $ and $ u_t $ as independent here.     

The portfolio allows short sales if $ \zeta_w > 1 $ with $ \zeta_w $ determining a constraint on the short sales as shown in Fan et al.\ (2012). Let 
$ w_0^{+} = \left( \zeta_w + 1 \right) / 2 $ and $ w_0^{-} = \left( \zeta_w - 1 \right) / 2$. Then $ w_0^{+} $ and  $ w_0^{-}$ correspond to the 
total proportions of long and short sales, respectively, since $ w_0^{+} + w_0^{-} = \zeta_w = \|\v{\omega}_0\|_1 $ and 
$ w_0^{+} - w_0^{-} = 1 $, and $ w_0^{-} $ becomes larger as $ \zeta_w $ grows while short sales are not allowed if $ \zeta_w = 1 $ ($ w_0^{-} = 0 $). We assume the fund manager holds equal amounts of long and short sales of $s/2 $ and that 
she employs equal weights among long and short sales; that is, we assume $ \omega_{0i} =  w_0^{+}/(s/2) $ for $ i \in \v{\omega}_{0A+}, $ $-w_0^{-}/(s/2) $ for $ i \in \v{\omega}_{0A-}, $ and  $ 0 $ for $ i \in \v{\omega}_{0B} $,  where $ \omega_{0i} $ 
is $i$th element of $ \v{\omega}_0 $, and $ \v{\omega}_{0A+} $, $ \v{\omega}_{0A-} $, and $ \v{\omega}_{0B} $ are sets of stocks of long, short, and no sales, respectively.

\subsubsection{Data and Evaluation Strategy}

We retrieve weekly stock price data of the  NYSE from {\it Yahoo! Finance}. Our dataset contains 1853 adjusted stock prices ($p=1853$) with starting from the 1st week of January in 2009 to the 4th week of April in 2016.
In this application, we apply the log-difference to the stock price data and standardize them so that the data are converted to rates of returns with zero means and unit variances. We investigate the cases of  $ s = 34 $ and $40$ with 
$ a = 14 $, $ \mathrm{SNR} = 10 $, and $ \zeta_w = 10 $. Non zero $s$ stocks are drawn randomly from $p$ candidates with equal probabilities. Furthermore, we assume the fund manager does not rebalance the portfolio. Hence it remain unchanged
 in all sample period. Brodie et al.\ (2009) argue a possibility of estimating a weight vector for a  portfolio in the presence of rebalancing with a penalized regression, but we do not consider the case here.

The purpose of this application is to screen  the kinds of stocks in which the fund manager invests from a large number of potential stocks using the penalized regression. We examine how well the penalized
 estimator $ \hat{\v{\omega}} $ can distinguish the nonzeros from zero elements of $ \v{\omega}_0 $ in finite samples. Then we evaluate the finite sample properties of 
$ \hat{\v{\omega}} $ to focus on  SC-$A$ $ = P\left( \mathrm{sgn}(\hat{\v{\omega}}_{A}) = \mathrm{sgn}(\v{\omega}_{0A} )\right)  $ and 
SC-$B$ $ = P\left( \mathrm{sgn}(\hat{\v{\omega}}_{B}) = \mathrm{sgn}(\v{\omega}_{0B} )\right)  $; the SC-$A$ is the success rate of detecting non-zero elements of $ \v{\omega}_0 $ with the correct sign and
 the SC-$B$ is that of detecting zero elements. We expect that the SCAD-type penalized regression estimator can have high SC-$A$ and SC-$B$ values as $T$ becomes large thanks to the oracle property.   
The SC-$A$ and SC-$B$ are sequentially computed for 172 evaluation periods in this application, where the endpoint gradually grows by one while the start point is fixed; the initial evaluation period starts 
from the 2nd week of January 2009 and ends in 1st week of December 2010 ($T=209$). The 2nd evaluation period runs from the 2nd week of January 2009 to the 2nd week of December 2010 ($T=210$), and so on. 
The terminal evaluation period is from the 2nd week of January 2009 to the 4th week of April  2016 ($T=381$).            

\subsubsection{Empirical Results} 

\begin{figure}[t!]
\begin{center}
\includegraphics[width=10cm,clip]{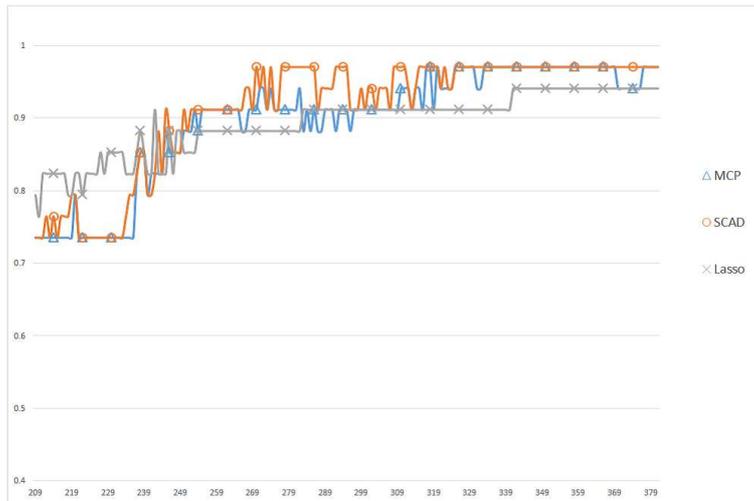}
\caption{SC-$A$ when $ s = 34 $ ~(from $ T = 209 $ to $ T = 381 $)}
\label{scA34}
\end{center}

\end{figure}

\begin{figure}[t!]
\begin{center}
\includegraphics[width=10cm,clip]{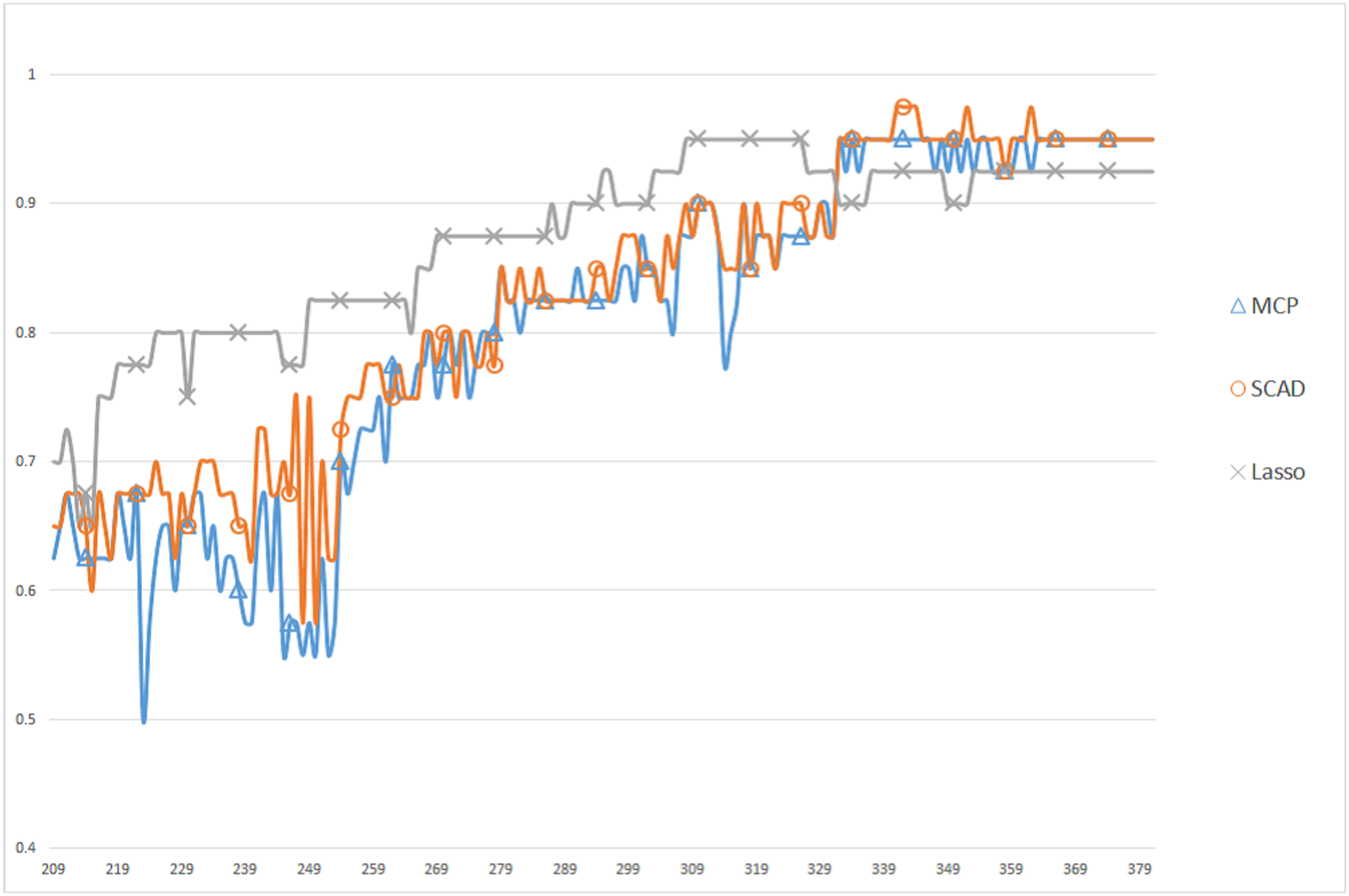}
\caption{SC-$A$ when $ s = 40 $ ~(from $ T = 209 $ to $ T = 381 $)}
\label{scA40}
\end{center}
\end{figure}

\begin{figure}[t!]
\begin{center}
\includegraphics[width=10cm,clip]{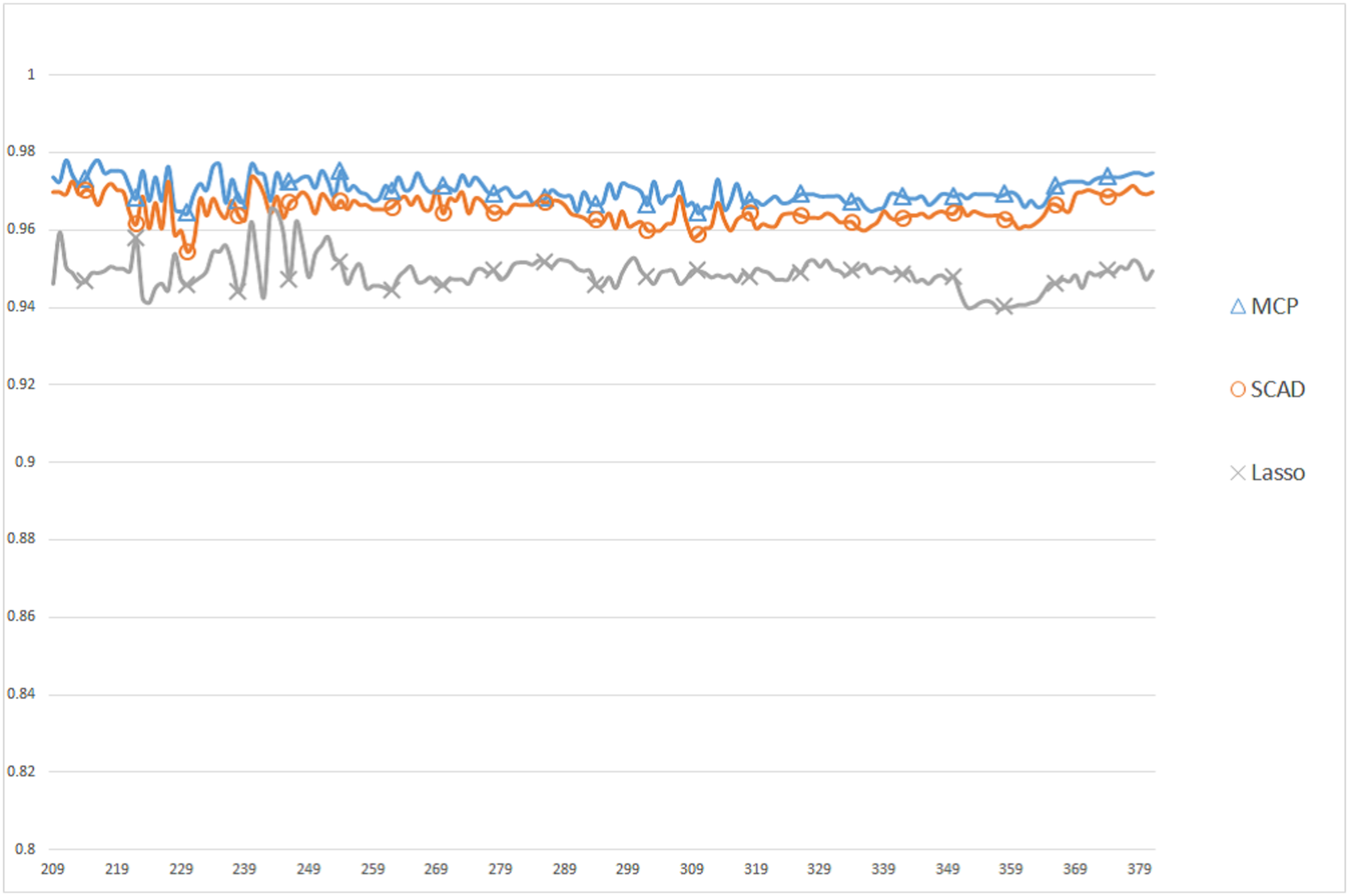}
\caption{SC-$B$ when $ s = 34 $ ~(from $ T = 209 $ to $ T = 381 $)}
\label{scB34}
\end{center}
\end{figure}

\begin{figure}[t!]
\begin{center}
\includegraphics[width=10cm,clip]{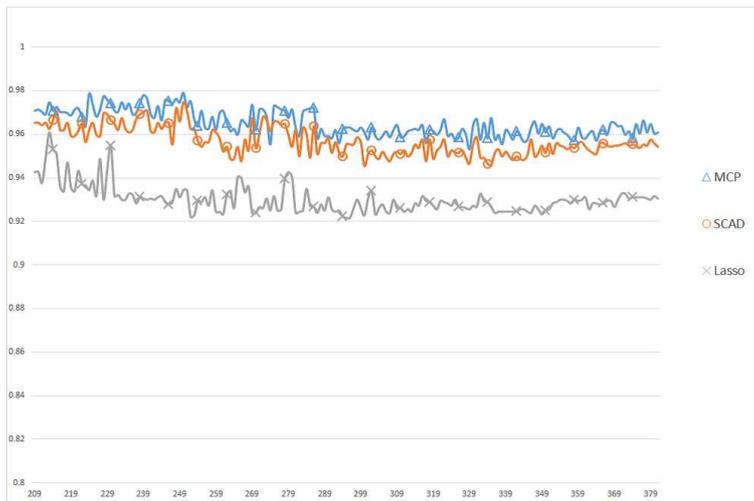}
\caption{SC-$B$ when $ s = 40 $ ~(from $ T = 209 $ to $ T = 381 $)}
\label{scB40}
\end{center}
\end{figure}

Figures \ref{scA34}--\ref{scA40} and Figures \ref{scB34}--\ref{scB40} show the SC-$A$ and SC-$B$ of the MCP, SCAD, and Lasso for 172 evaluation periods with $s=34 $ and $ 40 $, respectively.
To begin with, we consider the SC-$A$. At a glance, both Figures \ref{scA34} and \ref{scA40} reveal two characteristics of $ \v{\hat{\omega}} $. First, the SC-$A$ increases toward 1 as $ T $ grows 
for all penalties. Although the SC-$A$ of $ s=40$ seems uniformly lower than that of  
$ s = 34 $ for all $T$, this is due to the fact that more nonzero elements requires a greater search cost. Second, the SC-$A$ of the Lasso tends to be higher than that of the MCP and SCAD when $ T $ is relatively small, while it seems reversed 
when $ T $ grows large. This is consistent with the theory because the Lasso tends to have many ``false positive" estimates. That is, it overestimates the total number of nonzero elements since it rarely satisfies the assumptions for model selection 
consistency, while the MCP and SCAD satisfy these assumptions in many cases, as argued in Appendix \ref{noteonlasso}. Then, the SC-$A$ of the Lasso is not expected to be higher than that of the MCP and SCAD when $T$ is large. 

Next, we focus on the SC-$B$. Figures \ref{scB34} and \ref{scB40} show that SC-$B$ of the MCP and SCAD are successfully nearly equal to 1 and dominate that of the Lasso for
 all $T$. 
The results are consistent with the theory because the MCP and SCAD have the oracle property, which means they can detect true zero parameters more precisely than the Lasso can, except for extraordinary cases.                            

In summary, our empirical results reveal that the model selection consistency of the SCAD-type penalty works well in a large stock price dataset. This implies that the penalized regression enables us to effectively detect the behavior of fund managers 
from large financial datasets. 

\section{Conclusion}\label{sec6}

We have studied macroeconomic forecasting and variable selection using  a folded-concave penalized 
regression with a very large number of predictors. The contributions include both theoretical and empirical results. 
The first half of the paper developed the theory for a folded-concave penalized regression in ultrahigh dimensions 
when the model exhibits time series dependences. Specifically, we have proved the oracle inequality 
and the oracle property under appropriate conditions for macroeconomic time series. The latter half of the paper provided 
two empirical applications that motivated us to use the penalized regression for a large macroeconomic dataset. The first was  
the forecasting of quarterly U.S.\ real GDP with a large amount of monthly macroeconomic data taken from the FRED-MD through the MIDAS regression framework;
the forecasting model consisted of more than 1000 monthly predictors including lags  while the sample size was much smaller than the total number of predictors. 
The forecasting performance of the penalized regression is promising one compared to that of the factor MIDAS proposed by Marcellino and Schumacher (2010) and the state-space 
(nowcasting) model of Ba\'{n}bura and Modugno (2013). The second application screened a portfolio that contained about 40 stocks from  more than 1800 stocks using  NYSE stock price data. 
The oracle property ensured the variable selection consistency, that is, the penalized regression with the SCAD-type penalty could detect the portfolio from the data theoretically. In fact, we observed 
that the variable selection consistency worked properly when screening the portfolio. Our theoretical and empirical contributions are expected to introduce econometricians to the world of ultrahigh dimensional 
macroeconomic data.        

\vvsp 
\noindent
{\large \textbf{Acknowledgement}}  

\vvsp
\noindent
The authors are grateful for the invaluable comments of  the associate editor, the anonymous referees, Jinchi Lv, Ryo Okui, Yohei Yamamoto, and the participants 
of the econometric workshop in Kyoto University. All remaining errors are ours. The authors also thank Michelle Modugno for providing MATLAB codes used in Ba\'{n}bura and Modugno (2013), which 
were helpful when coding our R codes. Uematsu acknowledges the financial support from a Grant-in-Aid for JSPS Fellows No.26-1905. Tanaka acknowledges the financial supports from a JSPS Grant-in-Aid for 
Young Scientists (B) No.16K17100 and the Joint Usage and Research Center, Institute of Economic Research, Hitotsubashi University.

\newpage 
\setcounter{section}{1}

\renewcommand{\thesection}{\Alph{section}}
\renewcommand{\thesubsection}{\thesection.\arabic{subsection}}

\noindent
{\large \textbf{Appendix}}  

\subsection{Assumptions for the oracle property}\label{assoracleprop}

\begin{ass}\label{assevent2}\rm 
	There is a sequence $\lambda =o(1)$ such that $\| \v{G}_{0BT} \|_\infty \leq \lambda/2$ holds with high probability. 
\end{ass}

\begin{ass}\label{asspen}\rm 
	$(s/T)^{1/2}\ll \lambda \ll d =o(1)$ and $p_\lambda'(d)=0$ for a sufficiently large $T$.
\end{ass}

\begin{ass}\label{asslik0}\rm 
	For all $i$, $\max_t \E\left[ (x_{ti}u_t)^2 \right] <\infty$. 
\end{ass}


\begin{ass}\label{asslikLLN}\rm 
	There exists a constant $c_H$ such that the Hessian submatrix satisfies with high probability, $\min_{\v{v}\in \mathbb{R}^s} \v{v}^\top\v{H}_{AAT}\v{v}\geq c_H\|\v{v}\|_2^2$.
\end{ass}

\begin{ass}\label{asslik}\rm 
	$c_I \leq \Lambda_{\min}(\v{I}_{0AA}) \leq \Lambda_{\max}(\v{I}_{0AA}) \leq 1/c_I $ for a (small) constant $c_I>0$. 
\end{ass}

\begin{ass}\label{asslik_submat}\rm 
	$\|\v{H}_{BAT} \|_{2,\infty} \equiv \max_{\|\v{v}\|_2=1}\|\v{H}_{BAT} \v{v}\|_\infty = O_p(1)$.
\end{ass}

\begin{ass}\label{asslikasyn}\rm 
	$\E |\xi_t|^{2+\delta} \leq c_\xi $ for some constant $c_\xi>0$. 
\end{ass}

\subsection{Model selection inconsistency of Lasso}\label{noteonlasso}

As far as forecasting is concerned, Theorem \ref{oracleineq} shows that the resulting performance does not depend on the choice of penalties. 
However, if we wish to know what variables should be selected, the situation changes. 
We argue that a key assumption for model selection consistency for the $\ell_1$-penalty (Lasso) does not hold while a SCAD-type penalty does.

Zhao and Yu (2006) studied a concept called sign consistency defined by 
$P(\sgn(\hat{\v{\beta}})-\sgn(\v{\beta}_0))\rightarrow 1$, which is stronger than model selection consistency. 
Under a deterministic covariate assumption, they show that the {\it weak irrepresentable condition} 
\begin{align*}
\| \v{H}_{BAT} \v{H}_{AAT}^{-1} \sgn(\v{\beta}_{0A}) \|_\infty <1 
\end{align*}
is necessary for the sign consistency of Lasso. 
To establish the model selection consistency of Lasso, we usually need a stronger condition 
\begin{align*}
\| \v{H}_{BAT} \v{H}_{AAT}^{-1} \|_\infty \leq C 
\ \ \mbox{for some}\ \  C \in (0,1), 
\end{align*}
which was supposed by Fan and Lv (2011). It seems difficult to prove model selection consistency for the Lasso without this condition; 
however, the condition may be easily violated. Let $\v{x}_{i}$, $i\in B$, be a column vector of $\v{X}_B$. Then, the left-hand side of the bound is 
\begin{align*}
\| \v{H}_{BAT} \v{H}_{AAT}^{-1} \|_\infty 
= \max_{i \in B} \|(\v{X}_A^\top  \v{X}_A)^{-1} \v{X}_A^\top  \v{x}_i\|_1 
=:\max_{i \in B} \|\hat{\v{\pi}}_i\|_1, 
\end{align*}
where $\hat{\v{\pi}}_i \in \mathbb{R}^q$ is regarded as the OLS estimator of regression of an irrelevant variable $\v{x}_i$ on important variables $\v{X}_A$. 
Due to stationarity, this is $O_p(q)$ provided that the regularity conditions for an asymptotic theory are satisfied. 
Even when $q$ is finite, it is unrealistic for this value to be strictly bounded by one since 
macroeconomic data have cross-sectional dependence in general.  
When lagged variables are included in $\v{X}$, the condition becomes more tight because $A$ and $B$ may share the same variable. Violation of the condition 
would lead to a collapse of economic interpretation of estimated coefficients with the Lasso. 

\subsection{Lemmas for Theorems \ref{oracleineq}} \label{subsecevent}

The following lemmas were given by Loh and Wainwright (2015, Lemma 4(b) and Lemma 5), 
and are consequences of Assumption \ref{asspenalty}. 
They are used to fill the gap between the $\ell_1$-norm and SCAD-type penalties. 
The proofs are omitted.  

\begin{lem}\label{lemLWlem4b}
Under Assumotion \ref{asspenalty}, any vector $\v{\beta}\in \mathbb{R}^p$ satisfies 
\begin{align*}
\lambda \|\v{\beta}\|_1 \leq \|p_\lambda(\v{\beta})\|_1 + (\mu/2)\|\v{\beta}\|_2^2. 
\end{align*}
\end{lem}

\begin{lem}\label{lemLWlem5}
Under Assumotion \ref{asspenalty}, for any vector $\v{\beta}\in \mathbb{R}^p$ such that  
$\xi p_\lambda(\v{\beta}_0) - p_\lambda(\v{\beta})>0$ and $\xi\geq 1$, we have
\begin{align*}
\xi \|p_\lambda(\v{\beta}_0)\|_1 - \|p_\lambda(\v{\beta})\|_1 
\leq \xi \lambda \|\v{\beta}_A-\v{\beta}_{0A}\|_1 - \lambda \|\v{\beta}_B-\v{\beta}_{0B}\|_1. 
\end{align*}
\end{lem}

\subsection{Lemmas for Theorem \ref{exist}}
In Lemma \ref{lemcharact} below, let $\hat{A}:=\{ j\in\{1,\dots,p\} : \hat{\beta}_{j}\not=0 \}$, a set of indices corresponding to
all nonzero components of $\hat{\v{\beta}}$, and $\hat{\v{\beta}}_{\hat{A}}$ denote a subvector of $\hat{\v{\beta}}$
formed by its restriction to $\hat{A}$. The other symbols are defined analogously.
Let $\circ$ denote the Hadamard product. The sign function $\sgn(\cdot)$ is applied coordinate-wise.
Define 
\begin{align*}
G_{\hat{A}T}(\hat{\v{\beta}}) &= -T^{-1}\v{X}_{\hat{A}}^\top \v{y} +T^{-1} \v{X}_{\hat{A}}^\top \v{X}_{\hat{A}} \hat{\v{\beta}}_{\hat{A}}, \\
G_{\hat{B}T}(\hat{\v{\beta}}) &= -T^{-1}\v{X}_{\hat{B}}^\top \v{y} +T^{-1} \v{X}_{\hat{B}}^\top \v{X}_{\hat{A}} \hat{\v{\beta}}_{\hat{A}}.
\end{align*}
Define the {\it local concavity} at $\v{b} \in \mathbb{R}^r$ with $\|\v{b}\|_0=r$ as  
$\kappa_\lambda (\v{b}) = \max _{1\leq j \leq r} - p_\lambda'' (|b_j|)$. 

\begin{lem} \label{lemcharact} Suppose Assumption \ref{asspenalty} holds. 
Then $\hat{\v{\beta}}$ is a strict local minimizer of $Q_T(\v{\beta})$ in (\ref{obj}) if
\begin{align}
&G_{\hat{A}T}(\hat{\v{\beta}})+ p_\lambda'(\hat{\v{\beta}}_{\hat{A}}) \circ \sgn( {\hat{\v{\beta}}}_{\hat{A}} )=0, \label{lem11} \\
&\| G_{\hat{B}T}(\hat{\v{\beta}}) \|_\infty <p_\lambda'(0+), \label{lem12} \\
&\Lambda_{\min} (\v{H}_{\hat{A}\hat{A}T}) > \kappa_\lambda (\hat{\v{\beta}}_{\hat{A}}). \label{lem13}
\end{align}
Conversely, any local minimizer of $Q_T(\v{\beta})$ must satisfy (\ref{lem11}), (\ref{lem12}), and (\ref{lem13}) with strict inequalities 
replaced by nonstrict ones.  
\end{lem}

The proof was given by Lv and Fan (2009, Theorem 1). 
Consider the case where $\hat{\v{\beta}}_{\hat{A}} \in \mathcal{N}_0$. 
Under Assumption \ref{asspen}, it holds that 
$\sup_{\v{\beta}_{A} \in \mathcal{N}_0} \kappa_\lambda (\v{\beta}_{A})=0$ for sufficiently large $T$. 
Thus, condition (\ref{lem13}) is satisfied as long as $\Lambda_{\min} (\v{H}_{\hat{A}\hat{A}T})$ is bounded away from zero.

\subsection{Proofs of Theorems \ref{oracleineq} and \ref{exist}} \label{app_proof}

{\flushleft \bf Proof of Theorem \ref{oracleineq} }
Because $\hat{\v{\beta}}$ minimizes $Q_T(\v{\beta})$, we have 
\begin{align*}
(2T)^{-1}\| \v{y} - \v{X}\hat{\v{\beta}} \|_2^2 + \| p_\lambda(\hat{\v{\beta}}) \|_1 
\leq (2T)^{-1}\| \v{y} - \v{X}\v{\beta}_0 \|_2^2+ \| p_\lambda(\v{\beta}_0) \|_1. 
\end{align*}
By model (\ref{model2}) and Holder's inequality, this can be rewritten and bounded as 
\begin{align}
(2T)^{-1}\| \v{X}(\hat{\v{\beta}}-\v{\beta}_0) \|_2^2  
&\leq T^{-1}\v{u}^\top \v{X} (\hat{\v{\beta}}-\v{\beta}_0)+ \| p_\lambda(\v{\beta}_0) \|_1 
- \| p_\lambda(\hat{\v{\beta}}) \|_1 \notag\\
&\leq \| T^{-1}\v{X}^\top \v{u}\|_\infty  \|\hat{\v{\beta}}-\v{\beta}_0\|_1
+ \| p_\lambda(\v{\beta}_0) \|_1 
- \| p_\lambda(\hat{\v{\beta}}) \|_1. \label{basic1}
\end{align}
In what follows, we have only to work on event $\mathcal{E}_1$ defined in Assumption \ref{assevent}. 
On the event, we have 
$\| T^{-1}\v{X}^\top \v{u}\|_\infty \leq \lambda/2$, so that (\ref{basic1}) becomes 
\begin{align}
(2T)^{-1}\| \v{X}(\hat{\v{\beta}}-\v{\beta}_0) \|_2^2  
\leq 2^{-1}\lambda \|\hat{\v{\beta}}-\v{\beta}_0\|_1
+ \| p_\lambda(\v{\beta}_0) \|_1 
- \| p_\lambda(\hat{\v{\beta}}) \|_1. \label{basic2}
\end{align}
By Lemma \ref{lemLWlem4b}, the first term in the upper bound of (\ref{basic2}) is 
further bounded by 
\begin{align}
\lambda \|\hat{\v{\beta}}-\v{\beta}_0\|_1 
&\leq \| p_\lambda(\hat{\v{\beta}}-\v{\beta}_0) \|_1 
+ (\mu/2) \|\hat{\v{\beta}}-\v{\beta}_0\|_2^2 \notag \\
&\leq \| p_\lambda(\hat{\v{\beta}})\|_1 + \|p_\lambda(\v{\beta}_0)\|_1 
+ (\mu/2) \|\hat{\v{\beta}}-\v{\beta}_0\|_2^2, \label{basic3}
\end{align}
where the last inequality follows from the subadditivity implied by the concavity 
of the penalty function.
On the other hand, since $\|\hat{\v{\beta}}-\v{\beta}_0\|_0\leq \|\hat{\v{\beta}}\|_0 + \|\v{\beta}_0\|_0 \leq m$ 
holds on the assumed parameter space due to $\|\v{\beta}_0\|_0=s$, Assumption \ref{assRSC} yields the lower bound of (\ref{basic2}); 
that is, we have on $\mathcal{E}_2$ defined in Assumption \ref{assRSC}
\begin{align}
T^{-1}\| \v{X}(\hat{\v{\beta}}-\v{\beta}_0) \|_2^2  
\geq \gamma \| \hat{\v{\beta}}-\v{\beta}_0\|_2^2 . \label{basic4}
\end{align}
Therefore, combining (\ref{basic2}) with (\ref{basic3}) and (\ref{basic4}) gives 
\begin{align}
(\gamma-\mu/2) \| \hat{\v{\beta}}-\v{\beta}_0\|_2^2  
\leq 3 \| p_\lambda(\v{\beta}_0) \|_1 - \| p_\lambda(\hat{\v{\beta}}) \|_1. \label{basic5}
\end{align}
In particular, (\ref{basic5}) implies 
$3 \| p_\lambda(\v{\beta}_0) \|_1 - \| p_\lambda(\hat{\v{\beta}}) \|_1 \geq 0$, 
so that we can apply Lemma \ref{lemLWlem5} to the right-hand side of (\ref{basic5}) 
to obtain
\begin{align}
(\gamma-\mu/2) \| \hat{\v{\beta}}-\v{\beta}_0\|_2^2  
\leq 3 \lambda \|\hat{\v{\beta}}_A-\v{\beta}_{0A}\|_1 
- \lambda \|\hat{\v{\beta}}_B-\v{\beta}_{0B}\|_1. \label{basic6}
\end{align}
Ignoring the last term and the Cauchy-Schwarz inequality lead to 
\begin{align*}
(\gamma-\mu/2) \| \hat{\v{\beta}}-\v{\beta}_0\|_2^2  
\leq 3 \lambda \|\hat{\v{\beta}}_A-\v{\beta}_{0A}\|_1 
\leq 3 s^{1/2}\lambda \|\hat{\v{\beta}}_A-\v{\beta}_{0A}\|_2
\leq 3 s^{1/2}\lambda \|\hat{\v{\beta}}-\v{\beta}_{0}\|_2,
\end{align*}
which concludes the error bound in the $\ell_2$-norm 
\begin{align}
\| \hat{\v{\beta}}-\v{\beta}_0\|_2  
\leq \frac{6s^{1/2} \lambda}{2\gamma-\mu}. \label{basic7}
\end{align}

Using (\ref{basic7}), we can obtain the error bound in the $\ell_1$-norm as well. 
Since (\ref{basic6}) also implies that $\|\hat{\v{\beta}}_B-\v{\beta}_{0B}\|_1 
\leq 3 \|\hat{\v{\beta}}_A-\v{\beta}_{0A}\|_1$, we have 
\begin{align}
\|\hat{\v{\beta}}-\v{\beta}_{0}\|_1 
&= \|\hat{\v{\beta}}_A-\v{\beta}_{0A}\|_1 + \|\hat{\v{\beta}}_B-\v{\beta}_{0B}\|_1 \notag \\
&\leq 4 \|\hat{\v{\beta}}_A-\v{\beta}_{0A}\|_1 \leq 4s^{1/2}\|\hat{\v{\beta}}_A-\v{\beta}_{0A}\|_2 
\leq 4s^{1/2}\|\hat{\v{\beta}}-\v{\beta}_{0}\|_2 
\leq \frac{24 s \lambda}{2\gamma-\mu}. \label{basic8}
\end{align}

Finally, we derive the prediction error bound from (\ref{basic8}). 
The Mean value theorem, Assumption \ref{asspenalty}, and the triangle inequality give 
\begin{align*}
\| p_\lambda(\v{\beta}_0) \|_1 - \| p_\lambda(\hat{\v{\beta}}) \|_1 
&= \sum_{j=1}^p \left( |p_\lambda(\beta_{0j})| - |p_\lambda(\hat{\beta}_{j})| \right) 
= \sum_{j=1}^p p_\lambda'(b_j)\left( |\beta_{0j}| - |\hat{\beta}_{j}| \right) \\
&\leq p_\lambda'(0+)\sum_{j=1}^p \left| |\beta_{0j}| - |\hat{\beta}_{j}| \right| 
\leq \lambda \|\hat{\v{\beta}}-\v{\beta}_0\|_1, 
\end{align*}
where $b_j$ is a point between $|\beta_{0j}|$ and $|\hat{\beta}_{j}|$. 
Hence, using (\ref{basic2}), we obtain   
\begin{align}
T^{-1}\| \v{X}(\hat{\v{\beta}}-\v{\beta}_0) \|_2^2  
\leq 3 \lambda \|\hat{\v{\beta}}-\v{\beta}_0\|_1 
\leq \frac{72 s \lambda^2}{2\gamma-\mu}. \label{basic9}
\end{align}
Results (\ref{basic7})--(\ref{basic9}) hold with probability at least $1-O(p^{-c_1})-O(\exp(-c_2T)$ by Assumptions \ref{assevent} and \ref{assRSC}. $\square$


{\flushleft \bf Proof of Theorem \ref{exist} }
First, we show results (a) and (b) through the following steps. 

{\it Step 1.}
We consider $Q_T(\v{\beta})$ in the correctly constrained space
$\{\v{\beta} \in \mathbb{R}^p: \v{\beta}_{B}=\v{0} \in \mathbb{R}^{p-s} \}$, which is the $s$-dimensional subspace
$\{\v{\beta}_A \in \mathbb{R}^s \}$. The corresponding objective function is given by
\begin{align} \label{pr1}
Q_T(\v{\beta}_A,\v{0}) = (2T)^{-1} \| \v{y} - \v{X}_A \v{\beta}_A\|_2^2 + \|p_{\lambda}(\v{\beta}_A)\|_1.
\end{align}
We now show the existence of a strict local minimizer $\hat{\v{\beta}}_{0A}$ of $Q_T(\v{\beta}_A,\v{0})$ such that
$\|\hat{\v{\beta}}_{0A}-\v{\beta}_{0A}\|=O_p((s/T)^{1/2})$. To this end, it is sufficient to prove that,
for a large constant $C>0$, the event 
\begin{align} \label{pr12}
\mathcal{E}_Q = \left\{ \inf_{\|\v{v}\|_2 =C} Q_T(\v{\beta}_{0A}+\v{v}(s/T)^{1/2} ,\v{0}) > Q_T(\v{\beta}_{0A},\v{0}) \right\}
\end{align}
occurs with probability tending to one. This implies that, with high probability, 
there is a local minimizer $\hat{\v{\beta}}_{0A}$ of $Q_T(\v{\beta}_A,\v{0})$ 
in the ball $\mathcal{N}_C \equiv \{ \v{\beta}_A \in \mathbb{R}^s: \| \v{\beta}_A-\v{\beta}_{0A} \|_2 \leq C(s/T)^{1/2} \}$.

By the definition of the objective function, we have
\begin{align} 
R_T(\v{v}) :=&~ Q_T(\v{\beta}_{0A}+ \v{v}(s/T)^{1/2},\v{0}) - Q_T(\v{\beta}_{0A},\v{0})\notag\\ 
=&~ (s/T)^{1/2}\v{v}^\top \v{G}_{0AT} + (s/T)\v{v}^\top \v{H}_{AAT}\v{v} \label{Q-Q1}\\
& \qquad \qquad + \|p_\lambda(\v{\beta}_{0A}+ \v{v}(s/T)^{1/2})\|_1 - \|p_\lambda(\v{\beta}_{0A})\|_1. \label{Q-Q2}
\end{align}
First, we evaluate the two terms in (\ref{Q-Q2}). The Mean value theorem gives
\begin{align} \label{proof222}
\|p_\lambda(\v{\beta}_{0A}+ \v{v}(s/T)^{1/2})\|_1 - \|p_\lambda(\v{\beta}_{0A})\|_1&= \sum_{j \in A} p_\lambda'(|\beta_{0j}^{\star}|)
(|\beta_{0j}+v_j(s/T)^{1/2}|-|\beta_{0j}|) \notag \\
&\leq p_\lambda'(d)(s/T)^{1/2}\|\v{v}\|_1,
\end{align}
where $|\beta_{0j}^{\star}|$ lies between $|\beta_{0j}|$ and $|\beta_{0j}+v_j(s/T)^{1/2}|$, and the last inequality follows from
the monotonicity of $p_\lambda'(\cdot)$, $\min_{j \in A} |\beta_{0j}^{\star}| \geq d$, and the triangle inequality. 
Eventually, the last term is zero by Assumption \ref{asspen}. 
Next, we consider (\ref{Q-Q1}). 
Since martingale difference sequences are serially uncorrelated, Assumption \ref{asslik0} entails that
\begin{align*} 
\E \| \v{G}_{0AT} \|_2^2 &=T^{-2}\E [\v{u}^\top \v{X}_A\v{X}_A^\top \v{u}] 
=T^{-2}\sum_{j\in A}\E [\v{u}^\top \v{x}_j\v{x}_j^\top \v{u}] \\
&=T^{-2}\sum_{j\in A} \E \left[ \left( \sum_{t=1}^Tx_{tj}u_t \right)^2 \right] 
= T^{-2}\sum_{j\in A} \sum_{t=1}^T \E \left[ \left(x_{tj}u_t \right)^2 \right]
= O(s/T). 
\end{align*}
This together with the Markov inequality implies that $\| \v{G}_{0AT} \|_2$ is $O_p((s/T)^{1/2})$. 
Therefore, the Cauchy-Schwarz inequality yields
\begin{align*} 
(s/T)^{1/2} |\v{v}^\top \v{G}_{0AT}| \leq (s/T)^{1/2} \|\v{v}\|_2 \| \v{G}_{0AT}\|_2 = O_p(s/T)\|\v{v}\|_2.
\end{align*}
Whereas, by Assumption \ref{asslikLLN}, we get 
\begin{align} 
(s/T)\v{v}^\top  \v{H}_{AAT} \v{v} 
\geq (s/T) \Lambda_{\min}(\v{H}_{AAT})\|\v{v}\|_2^2
\geq (s/T) c_H\|\v{v}\|_2^2. \label{hhh}
\end{align}
Because (\ref{hhh}) dominates the other terms of $R_T(\v{v})$ when a large value of $\|\v{v}\|_2$ is taken, 
$\inf_{\|\v{v}\|_2=C} R_T(\v{v})$ tends to positivity as $T$ grows large. Thus, with probability approaching one, 
(\ref{pr12}) holds, and $ \| \hat{\v{\beta}}_{0A}-\v{\beta}_{0A} \|_2 \leq C(s/T)^{1/2}$.

{\it Step 2.}
To complete the proof of (a) and (b), it remains to show that $\hat{\v{\beta}}_0:=(\hat{\v{\beta}}_{0A},\v{0})$ is
indeed a strict local maximizer of $Q_T(\v{\beta})$ in $\mathbb{R}^p$. From Lemma \ref{lemcharact}, 
it suffices to check conditions (\ref{lem11}), (\ref{lem12}), and (\ref{lem13}) 
with setting $\hat{\v{\beta}}=\hat{\v{\beta}}_0$, but 
condition (\ref{lem11}) is satisfied by the proof of Theorem 1 in Fan and Lv (2011). 

We then check Condition (\ref{lem13}). Define $\mathcal{N}_0:=\{\v{\beta}_A \in \mathbb{R}^s : \|\v{\beta}_A-\v{\beta}_{0A}\|_\infty \leq d\}$, where we recall $d = \min_{j \in A}|\beta_{0,j}|/2$.  
By Assumption \ref{asspen}, we have $d/(s/T)^{1/2} \rightarrow \infty$, so that, for sufficiently large $T$, $\hat{\v{\beta}}_A \in \mathcal{N}_C$ 
implies $\hat{\v{\beta}}_A \in \mathcal{N}_0$. 
Thus the condition is eventually satisfied by Assumptions  \ref{asslikLLN} and \ref{asslik} along with the comment after Lemma \ref{lemcharact}.

To verify (\ref{lem12}), we first see that $(s/T)^{1/2}/\lambda=o(1)$ by Assumption \ref{asspen}. Thus, Assumptions \ref{assevent2} and \ref{asslik_submat} establish 
\begin{align*} 
\|G_{BT}(\hat{\v{\beta}})\|_\infty 
&= \| \v{H}_{BAT} ( \hat{\v{\beta}}_{A} - \v{\beta}_{0A} ) + \v{G}_{0BT} \|_\infty 
\leq \| \v{H}_{BAT} ( \hat{\v{\beta}}_{A} - \v{\beta}_{0A} ) \|_\infty + \| \v{G}_{0BT} \|_\infty \notag \\
&\leq \| \v{H}_{BAT}\|_{2,\infty} \| \hat{\v{\beta}}_A-\v{\beta}_{0A} \|_2 + \lambda/2 \\
&= O_p(1)C(s/T)^{1/2} + \lambda/2 = \left\{ o_p(1)+1 \right\}\lambda/2.
\end{align*}
Since $p_\lambda'(0+)=\lambda$ in Assumption \ref{asspenalty}, condition (\ref{lem12}) holds for a sufficiently large $T$. This completes the proof of (a) and (b).

Finally, we prove (c). Clearly we only need to show the asymptotic normality of $\hat{\v{\beta}}_A$. 
Assumption \ref{asslik} ensures that $\v{I}_{0AA}$ is positive definite, and hence, $\v{I}_{0AA}^{-1/2}$ is well-defined. On the event $\mathcal{E}_Q$ in (\ref{pr12}), it has been shown that $\hat{\v{\beta}}_A \in \mathcal{N}_C$ is a strict local minimizer 
of $Q_T(\v{\beta}_A,\v{0})$ and $\partial Q_T(\hat{\v{\beta}}_A,\v{0})/\partial \v{\beta}_A= \v{0}$. 
We thus obtain, for any vector $\v{b}\in \mathbb{R}^s$ such that $\|\v{b}\|_2=1$, 
\begin{align} 
-T^{1/2}\v{b}^\top \v{I}_{0AA}^{-1/2} \v{H}_{AAT}(\hat{\v{\beta}}_A-\v{\beta}_{0A}) = T^{1/2}\v{b}^\top \v{I}_{0AA}^{-1/2} \v{G}_{0AT} + T^{1/2}\v{b}^\top \v{I}_{0AA}^{-1/2} p_\lambda'(\hat{\v{\beta}}_A)\circ \sgn(\hat{\v{\beta}}_A). \label{focasyn} 
\end{align}
Recall that $T^{1/2}\v{b}^\top \v{I}_{0AA}^{-1/2} \v{G}_{0AT}=\sum_{t=1}^T\xi_{Tt}$ and $\xi_{Tt}$ is a martingale difference array. 
We show the asymptotic normality of this part. It is not hard to say that
\begin{align*} 
\sum_{t=1}^T {\rm Var} (\xi_{Tt}) =\v{b}^\top \v{I}_{0AA}^{-1/2}\v{I}_{0AA} \v{I}_{0AA}^{-1/2}\v{b} =1. 
\end{align*}
Assumption \ref{asslikasyn} implies uniform integrability of $\xi_t^2$. Hence, by Theorems 24.3 and 24.4 of Davidson (1994, Ch.\ 24), 
we obtain $\sum_{t=1}^T \xi_{Tt} \rightarrow_d N(0,1)$. 
Because the last term of (\ref{focasyn}) is $o_p(1)$ by the argument above, 
the result follows from the Slutsky lemma and Assumption \ref{asslikLLN}. $\square$

\subsection{ Lemmas for Proposition \ref{propassevent}} \label{lemproofpropassevent}

Recall that $c_{xu} = \limsup_T \max_{t,i} \{R_{tT}^{(X)}\Sigma_{pi}, R_{tT}^{(u)}\Sigma_{pi}\} <\infty$.

\begin{lem} \label{lemxuboundu} Under Assumption \ref{covariate1}, we have for any $i$ and $\alpha>0$,
\begin{align*}
P \left( \max_t \left| x_{ti}  u_t \right| > \alpha  \right) &\leq 4T \exp \left\{ -\alpha /(2 c_{xu}) \right\}. 
\end{align*} 
\end{lem}

{\flushleft \bf Proof } We see that
\begin{align*}
P \left( \left| x_{ti}  u_t \right| > \alpha  \right) 
&\leq P \left(  | x_{ti} |  > \alpha^{1/2} \right) + P \left( | u_t |  > \alpha^{1/2}  \right).
\end{align*}
We consider the first term. By the construction of $x_{ti}$ with suppressing the superscript, we have
\begin{align*}
P \left( \left| \sum_{s=1}^T \sum_{k=1}^p r_{ts}\sigma_{ki} z_{sk}  \right|  > \alpha^{1/2} \right) 
&= P \left( (R_{tT}\Sigma_{pi})^{-1/2}\left| \sum_{s=1}^T \sum_{k=1}^p r_{ts}\sigma_{ki} z_{sk}  \right|  > (\alpha/R_{tT}\Sigma_{pi})^{1/2} \right) \\
&\leq 2 \exp \left\{ -\alpha/(2R_{tT}\Sigma_{pi}) \right\}
\leq 2 \exp \left\{ -\alpha/(2c_{xu}) \right\}, 
\end{align*}
where the first inequality holds since $(R_{tT}\Sigma_{pi})^{-1/2} \sum_{s=1}^T \sum_{k=1}^p r_{ts}\sigma_{ki} z_{sk}$ is a standard normal random variable and the last inequality follows from Assumption \ref{covariate1}. 
It is clear that we obtain the same result for $u_t$. Therefore, by the union bound, we have 
\begin{align*}
P \left( \max_t \left| x_{ti}  u_t \right| > \alpha  \right) 
\leq 4T \exp \left\{ -\alpha/(2c_{xu}) \right\}, 
\end{align*} 
which yields the desired inequality. $\square$

\begin{lem} \label{sumxuAzuma} Let $ \lambda = c_0\log(pT) (\log p/T)^{1/2} $ for any positive constants $ c_0 $. Let $m_0$ be an arbitrary positive constant.  
Under Assumption \ref{covariate1}, we have for any $i$, 
\begin{align*}
P \left(T^{-1}|\v{x}_i^\top \v{u}| \geq \lambda/2 ~ \big|  \max_t | x_{ti} u_t | \leq m_0 \log (pT) \right) \leq  2 p^{- c_0^2/(8m_0^2)}.
\end{align*} 
\end{lem}

{\flushleft \bf Proof } 
Because $ \left( x_{ti} u_t, \mathcal{F}_t \right) $ is a martingale difference sequence with respect to $  \mathcal{F}_t = \{ u_{t-j}, x_{t-j+1}:j=0,1,\dots \} $ for each $i$, Azuma-Hoeffding's inequality yields 
\begin{align*}
P \left(T^{-1}|\v{x}_i^\top \v{u}| \geq \lambda/2 ~ \big|  \max_t | x_{ti} u_t | \leq \alpha \right)  &=  P \left( \left| \sum_{t=1}^T x_{ti} u_t \right| \geq T \lambda/2 ~ \big|  \max_t | x_{ti} u_t | \leq \alpha \right)  \\ 
 &\leq 2 \exp \left[ - \dfrac{(T\lambda/2)^2}{2T \alpha^2} \right]  
\end{align*}
for any $\alpha>0$ for each $T$ and $p$. Plugging $\lambda$ and $\alpha=m_0 \log (pT)$ into the upper bound, we have 
\begin{align*}
2 \exp \left[ - \dfrac{ T c_0^2 (\log pT)^{2} \log p}{8T m_0^2 (\log pT)^{2}} \right] 
= 2 \exp \left[ - \dfrac{ c_0^2 \log p}{8 m_0^2} \right] 
= 2 p^{-c_0^2/(8m_0^2)},
\end{align*}   
giving the result. $\square$

\subsection{ Proofs  of Propositions \ref{propassevent} and \ref{propassRSC}} \label{proofpropassevent}

{\flushleft \bf Proof of Proposition \ref{propassevent} } 
By the union bound and the property of the conditional probability, we have for any $\alpha>0$,
\begin{align*}
P(\mathcal{E}_1^c)   &= P( T^{-1} \max_i | \v{x}_i^{\top} \v{u} | \geq \lambda/2 ) 
                         \leq  \sum_{i=1}^p P( T^{-1} | \v{x}_i^{\top} \v{u} | \geq \lambda/2 ) \\ 
                         &\leq \sum_{i=1}^p P( T^{-1} | \v{x}_i^{\top} \v{u} | \geq \lambda/2  ~|  \max_t | x_{ti} u_t | \leq \alpha )  + \sum_{i=1}^p P( \max_t | x_{ti} u_t | > \alpha).
\end{align*}
Let $ \alpha = m_0 \log (pT) $ be the same as in the proof of Lemma \ref{sumxuAzuma}. 
From Lemmas \ref{lemxuboundu} and \ref{sumxuAzuma}, this is bounded as 
\begin{align*}
P(\mathcal{E}_1^c) 
&\leq 2 p^{-\left\{c_0^2/(8m_0^2)-1 \right\}} + 4 pT \exp \left\{ -m_0 \log (pT)/(2 c_{xu})  \right\} \\
&\leq 2 p^{-c_0^2/(8m_0^2)+1 } + 4 (pT)^{-m_0/(2 c_{xu}) + 1 }.
\end{align*}
Since $m_0$ and $c_0$ are arbitrary, putting $m_0=c_0/4$ and $c_0\geq 16 c_{xu}$ reduces to 
\begin{align*}
P(\mathcal{E}_1^c) 
&\leq 2p^{-1} + 4(pT)^{-c_0/(8c_{xu})+1} \leq 2p^{-1} + 4(pT)^{-1} \leq 6p^{-1},
\end{align*}
giving the result. $\square$

{\flushleft \bf Proof of Proposition \ref{propassRSC} } 
We have 
\begin{align}
m \leq \phi T^{1-\delta}<T,\label{ineq:d}
\end{align}
where the last strict inequality holds for large $T$. 
Note that each row of $\v{W} \equiv \v{Z} \v{\Sigma}_X^{1/2}$ is viewed as an $p$-dimensional random vector independently sampled from $N(0,\v{\Sigma}_X)$. Let $\tilde{\v{d}}=\v{d}_{\supp(d)}$, 
$\tilde{\v{X}}=\v{X}_{\supp(d)}$ and $\tilde{\v{W}}=\v{W}_{\supp(d)}$. 
For any $\supp(\v{d})\subset \{1,\dots,p\}$ satisfying (\ref{ineq:d}) and $ \| \v{d} \|_0 \leq m $, we see that 
\begin{align}
T^{-1}\|\tilde{\v{X}}\tilde{\v{d}}\|_2^2/\|\tilde{\v{d}}\|_2^2
&= T^{-1} 
\left(\frac{\tilde{\v{d}}^\top \tilde{\v{W}}^\top \v{R}_X \tilde{\v{W}} \tilde{\v{d}}}{\tilde{\v{d}}^\top \tilde{\v{W}}^\top \tilde{\v{W}} \tilde{\v{d}}} \right) 
\left( \frac{\tilde{\v{d}}^\top \tilde{\v{W}}^\top \tilde{\v{W}} \tilde{\v{d}}}{\tilde{\v{d}}^\top  \tilde{\v{d}}} \right)\notag\\
&\geq T^{-1} \min_{\v{h}\in \mathbb{R}^{T}}\left( \frac{\v{h}^\top \v{R}_X \v{h}}{\v{h}^\top \v{h}} \right)
\min_{\tilde{\v{d}}\in\mathbb{R}^m}\left( \frac{\tilde{\v{d}}^\top \tilde{\v{W}}^\top \tilde{\v{W}} \tilde{\v{d}}}{\tilde{\v{d}}^\top  \tilde{\v{d}}} \right)\notag\\
&\geq c_R \min_{\tilde{\v{d}}\in\mathbb{R}^m}T^{-1} \|\tilde{\v{W}} \tilde{\v{d}}\|_2^2/\|\tilde{\v{d}}\|_2^2,\label{ineq:quadform}
\end{align}
where the last inequalities hold by Assumption \ref{covariate1}. 
We denote by $\tilde{\v{w}}_t$ and $\tilde{c}$ the $t$th row of $\tilde{\v{W}}$ and the minimum eigenvalue of the covariance matrix of $\tilde{\v{w}}_t$, respectively. 
Inequality (\ref{ineq:quadform}) with the fact that $\tilde{c}\geq c_{\Sigma}$ leads to 
\begin{align}
P\left( \min_{\tilde{\v{d}} \in \mathbb{R}^m} T^{-1}\|\tilde{\v{X}} \tilde{\v{d}}\|_2^2/\|\tilde{\v{d}}\|_2^2 \leq c_{\Sigma} c_R/9 \right)
&\leq P\left( c_R\min_{\tilde{\v{d}} \in \mathbb{R}^m} T^{-1}\|\tilde{\v{W}} \tilde{\v{d}}\|_2^2/\|\tilde{\v{d}}\|_2^2 \leq c_{\Sigma} c_R/9 \right) \notag\\
&\leq P\left( \min_{\tilde{\v{d}} \in \mathbb{R}^m} T^{-1}\|\tilde{\v{W}} \tilde{\v{d}}\|_2^2/\|\tilde{\v{d}}\|_2^2 \leq \tilde{c}/9 \right). \label{ineq:XW}
\end{align}
An application of Lemma 9 in Wainwright (2009) gives 
\begin{align}
P\left( \min_{\tilde{\v{d}} \in \mathbb{R}^m} T^{-1}\|\tilde{\v{W}} \tilde{\v{d}}\|_2^2/\|\tilde{\v{d}}\|_2^2 \leq \tilde{c}/9 \right) \leq 2\exp(-T/2).\label{ineq:GaussConcent}
\end{align}

Finally, we extend the result uniformly in terms of the choice of $\supp(\v{d})$. We see that $\binom pm \leq p^m \leq \exp(\phi^2 T)$ holds for large $T$ by Stirling's approximation and (\ref{ineq:d}) with Assumption \ref{assdim}. Therefore, taking the union bound with combining (\ref{ineq:XW}) and (\ref{ineq:GaussConcent}) gives
\begin{align*}
P\left( \min_{\v{d}\in \mathbb{R}^p, \, \|\v{d}\|_0\leq m} T^{-1}\|\v{X} \v{d}\|_2^2/\|\v{d}\|_2^2 \leq c_Gc_R/9 \right)
\leq 2\exp(\phi^2 T-T/2), 
\end{align*}
which goes to zero since $\phi^2<1/2$ by Assumption \ref{assdim}. 
Consequently, if we choose $\gamma = c_Gc_R/9$ and $c_2 = 1/2-\phi^2$, we achieve the result.  $\square$

\subsection{Collinearity}\label{subseccollinear}


We explore how collinearity between $\v{X}_B$ and $\v{X}_A$ affects the oracle property 
obtained by Theorem \ref{exist}. Assumption \ref{asslik_submat} controls how much collinearity is allowed. 
Recall that $\v{H}_{BAT}=T^{-1}\v{X}_B^\top \v{X}_A$ for $\v{X}_A\in \mathbb{R}^{T\times s}$ 
and $\v{X}_B\in \mathbb{R}^{T\times (p-s)}$. We are interested in the behavior of  
\begin{align*}
\|\v{H}_{BAT}\|_{2,\infty} &\equiv \max_{\|\v{v}\|_2=1} \|\v{H}_{BAT}\v{v}\|_\infty
= \max_{b \in B} \max_{\|\v{v}\|_2=1} \left|T^{-1} \v{x}_b^\top \v{X}_A \v{v}\right|, 
\end{align*}
where we write $\v{X}_A \v{v} = \sum_{a \in A} v_a\v{x}_a$. 
This value is expected to become unbounded (and hence Assumption \ref{asslik_submat} is violated) 
under strong collinearity.

To obtain understandable results, we make the following simplified assumptions: the regressors are deterministic, 
and for any $b \in B$ and $a \in A$, $T^{-1}\v{x}_b^\top \v{x}_a \to \rho_{ba} \geq 0$. Moreover, 
we assume either of the two conditions: 
\begin{enumerate}
\item $\max_{b\in B}\rho_{ba} \geq c>0$ for all $a\in A$, 
\item $\max_{b\in B}\rho_{ba} \leq ca^{-q/2}$ for some $q > 1$.
\end{enumerate}
Condition 1 describes a highly correlated case. The correlation between $\v{x}_b$ and $\v{x}_a$ always exists 
even if $s$ increases. 
On the other hand, condition 2 models weaker correlations than condition 1 does. Specifically, most of the correlations 
become small as $q$ becomes large, meaning that the effect of collinearity is limited in this case. 
In fact, it is not difficult to see that 
$\|\v{H}_{BAT}\|_{2,\infty}$ diverges at least as fast as $s^{1/2}$ under condition 1 while  
$\|\v{H}_{BAT}\|_{2,\infty}$ is uniformly bounded under condition 2:
First, we suppose condition 1 and let $\bar{\v{v}}=(s^{-1/2},\dots,s^{-1/2})^\top$. We then observe that 
\begin{align*}
\max_{b \in B} \max_{\|\v{v}\|_2=1} \left|T^{-1} \v{x}_b^\top \v{X}_A \v{v}\right| 
\geq \max_{b \in B} \left|T^{-1} \v{x}_b^\top \v{X}_A \bar{\v{v}} \right| 
= \max_{b \in B} \left|s^{-1/2}\sum_{a \in A} T^{-1} \v{x}_b^\top \v{x}_a\right|. 
\end{align*}
By condition 1, the last term is bounded from below by 
\begin{align*}
s^{-1/2}\max_{b \in B} \left|\sum_{a \in A} \left( \rho_{ba}+o(1) \right) \right| 
\geq s^{1/2}(c - o(1)), 
\end{align*}
which goes to infinity as $s\to \infty$. 
Next, we suppose condition 2. By the Cauchy-Schwarz inequality, we observe that 
\begin{align*}
\max_{b \in B} \max_{\|\v{v}\|_2=1} \left|T^{-1} \v{x}_b^\top \v{X}_A \v{v}\right| 
&= \max_{b \in B} \max_{\|\v{v}\|_2=1} \left|\sum_{a \in A} v_a T^{-1} \v{x}_b^\top \v{x}_a \right| \\
&= \max_{b \in B} \max_{\|\v{v}\|_2=1} \left|\sum_{a \in A} v_a \left( \rho_{ba}+o(1) \right) \right| \\
&\leq \max_{b \in B} \left( \sum_{a \in A} \rho_{ba}^2\left( 1+o(1) \right) \right)^{1/2} 
\leq c\left( \sum_{a \in A} a^{-q}\left( 1+o(1) \right) \right)^{1/2}. 
\end{align*}
The last term converges since $q>1$ under condition 2.

The following simulation shows that the strong collinearity (condition 1) affects the oracle property. 
Table \ref{collinear} shows the {\it relative} finite sample success rates detecting non-zero ($SC$-$A$) coefficients and zero coefficients ($SC$-$B$) that are defined as $SC$-$A$ $ =  P\left( \mathrm{sgn}(\hat{\v{\beta}}_{A}) = \mathrm{sgn}(\v{\beta}_{0A} )\right)  $ 
and SC-$B$ $ = P\left( \mathrm{sgn}(\hat{\v{\beta}}_{B}) = \mathrm{sgn}(\v{\beta}_{0B} )\right)  $ respectively, and (average) mean squared error for estimates of non-zero coefficients ($ \mathrm{MSE} (\hat{\v{\beta}}_A) $)  under Condition 1 compared to that of 
Condition 2 when $ T = 300, 500, 1000$ and $ c = 0.5, 0.98 $ with  
$ q = 4 $, $ p = 1.5 \exp(T^{0.31}) $ and $ s = 20 T^{0.3} $. Then, the finite sample properties of estimators under Condition 1 are equivalent to those of Condition 2 if  the values in the Table are 1.  We can confirm  facts  
from Table \ref{collinear} that ($i$) the success rates are relatively low under Condition 1 irrespective of  the degree of collinearity ($c$)  and ($ii$) the $MSE$ of the Condition 1 is expected to be much worse than the that of Condition 2 
asymptotically especially when the degree of collinearity is high. These facts are consistent to the theoretical results because the Condition 1 violates  Assumption \ref{asslik_submat} so that the oracle property no longer holds under Condition 1.

\begin{table}[t!]
\caption{Relative $ SC$-$A$, $ SC$-$B $ and $ MSE $ (cond.1/cond.2)}
\begin{center}
\begin{tabular}{rrrrrrrr} 
 & \multicolumn{3}{r}{$ c = 0.5 $} &  & \multicolumn{3}{r}{$ c = 0.98 $} \\
 & $ SC-A $ & $ SC-B $ & $ MSE $ &  & $ SC-A $ & $ SC-B $ & $ MSE $ \\\cline{2-4}\cline{6-8}
$ T = 300 $ & 0.89 & 0.98 & 1.10 &  & 0.96 & 1.01 & 0.99 \\
$ T = 500 $ & 0.88 & 0.99 & 1.19 &  & 0.96 & 1.00 & 0.98 \\
$ T = 1000$  & 1.00 & 1.00 & 1.25 &  & 0.95 & 1.00 & 3.49 \\
\end{tabular}
\end{center}
\label{collinear}
\end{table}

\subsection{Related works}\label{subsecrelation}


Wang et al.\ (2007) investigated the asymptotic properties of the Lasso and modified  Lasso (Lasso$^*$) for the linear regression with the autoregressive error model. They derived the model selection consistency, and showed the Lasso$^*$ can be the oracle estimator. 
Nardi and Rinaldo (2011) considered the estimation and variable (lag) selection of autoregressive models via the Lasso. They mainly focused on the lag selection of the AR parameters. 
Lasso-type estimation of VAR models has been studied by several authors, including Song and Bickel (2011), Nicolson et al.\ (2015), Basu and Michailidis (2015), and Kock and Callot (2015). Theoretically, the latter two papers have significant contribution to the high-dimensional time series literature, but their settings are different from ours. The results obtained here are new and much complement their works. 
Basu and Michailidis (2015) investigated estimation of general high-dimensional time dependent models via spectral densities of covariates and errors, and derived the non-asymptotic error bound. 
Kock and Callot (2015) derived the non-asymptotic error bound for a high-dimensional VAR model.

\end{document}